\documentclass[review]{elsarticle}

\journal{Physics of the Earth and Planetary Interiors}

\biboptions{authoryear}

\usepackage{subcaption}
\usepackage{mathtools}
\graphicspath{{fig/}}
\usepackage{multirow}
\usepackage{rotating}
\usepackage{booktabs,array}
\usepackage{xspace}

\usepackage{soul}
\DeclareRobustCommand{\soulwhite}[1]{{\sethlcolor{white}\hl{#1}}}

\newcommand{\Acronym}{CPIC\xspace}

\newcommand{\MW}[1]{M\textsubscript{W}\,#1}
\newcommand{\ML}[1]{ML\,#1}
\begin{document}

\begin{frontmatter}
    
    %\title{Elsevier \LaTeX\ template\tnoteref{mytitlenote}}
    \title{Deep learning for seismic phase detection and picking in the aftershock zone of 2008 $M_w 7.9$ Wenchuan Earthquake}
    
    
    
    \author[ece]{Lijun Zhu\corref{ca}}
    \cortext[ca]{Corresponding author.}
    \ead{lijun.zhu@gatech.edu}

    \author[eas]{Zhigang Peng}
    
    \author[ece]{James McClellan}
    
    \author[eas]{Chenyu Li}
    
    \author[eas]{DongDong Yao}
    
    \author[catech]{Zefeng Li}
    
    \author[tmp]{Lihua Fang}
    
    \address[ece]{School of Electrical and Computer Engineering\\
        Georgia Institute of Technology\\
        Atlanta, GA 30332, U.S.A.}
    \address[eas]{School of Earth and Atmospheric Sciences\\
        Georgia Institute of Technology\\
        Atlanta, GA 30332, U.S.A.}
    \address[catech]{Seismological Laboratory\\
        Division of Geological and Planetary Sciences\\
        California Institute of Technology\\
        Pasadena, CA 91125, U.S.A.}
    
    \address[tmp]{Institute of Geophysics\\
        China Earthquake Administration\\
        Beijing, 100081, China}
    
    \begin{abstract}
        The increasing volume of seismic data from long-term continuous monitoring motivates the development of algorithms based on convolutional neural network (CNN) for faster and more reliable phase detection and picking.
        However, many less studied regions lack a significant amount of labeled events needed for traditional CNN approaches.
        In this paper, we present a CNN-based Phase-Identification Classifier (\Acronym) designed for phase detection and picking on small to medium sized training datasets.
        When trained on 30,146 labeled phases and applied to one-month of continuous recordings during the aftershock sequences of the 2008 \MW{7.9} Wenchuan Earthquake in Sichuan, China, \Acronym detects 97.5\% of the manually picked phases in the standard catalog and predicts their arrival times with a five-times improvement over the ObsPy AR picker.
        In addition, unlike other CNN-based approaches that require millions of training samples, when the off-line training set size of \Acronym is reduced to only a few thousand training samples the accuracy stays above 95\%.
        The online implementation of \Acronym takes less than 12 hours to pick arrivals in 31-day recordings on 14 stations.
        In addition to the catalog phases manually picked by analysts, \Acronym finds more phases for existing events and new events missed in the catalog.
        Among those additional detections, some are confirmed by a matched filter method while others require further investigation.
        Finally, when tested on a small dataset from a different region (Oklahoma, US), \Acronym achieves 97\% accuracy after fine tuning only the fully connected layer of the model.
        This result suggests that the \Acronym developed in this study can be used to identify and pick P/S arrivals in other regions with no or minimum labeled phases.
    \end{abstract}
    
    
\end{frontmatter}

%\linenumbers

\section{Introduction}
%%% Intro to Phase Picking %%%
Event detection and phase picking algorithms are becoming increasingly important for automatic processing of large seismic datasets.
Reliable automatic methods for P-wave picking have been available for decades.
The commonly adopted approaches for automatic picking of seismic phases convert the time-domain signal to a characteristic function (CF), such as short-term/long-term average (STA/LTA) \citep{Allen1982}, envelope functions \citep{Baer1987}, or autoregressive modeling of Akaike Information Criterion (AR-AIC) \citep{Sleeman1999}, and then select the indices of local maxima, or their rising edges, as the picked arrival times.
Higher-order statistics, including kurtosis \citep{Saragiotis2002} and skewness \citep{Nippress2010, Ross2014b}, have also been used to refine the picks due to their sensitivity to abrupt changes in a time series.
These algorithms generally perform better for the P waves than S waves, most likely because S-wave arrivals are usually contaminated by the P coda and converted phases.
Polarization has been used to discriminate P and S phases \citep{Jurkevics1988}.
The covariance matrix \citep{Cichowicz1993} is used to rotate waveforms into polarized P and S waveform components using methods such as singular value decomposition (SVD) \citep{Rosenberger2010,Kurzon2014}.
In general, these existing methods make certain assumptions about the observed seismograms and require careful parameter tweaking when operating on different datasets.

%%% Waveform similarity approaches %%%
Recently, waveform similarity has been used to detect earthquakes originating from a small region with the same source mechanism while using relatively few parameters \citep{Gibbons2006,Shelly2007,Peng2009}.
% Template matching
A subset of the events with high signal-to-noise ratio (SNR) is manually picked as templates to cross-correlate with continuous waveforms to detect smaller events similar to these templates.
The computation cost of such template matching methods scales linearly with respect to the number of templates and dataset size.
Since the detected events must be similar to one of the template events, this approach is not as general as the aforementioned STA/LTA.
% Autocorrelation
Waveform \textit{autocorrelation} is one of the most effective methods to detect nearly repeating seismic signals \citep{Brown2008}.
Despite being reliable and robust for different regions, its computation cost scales quadratically with the size of the dataset, making it infeasible when scaled to longer time periods.
Further efforts have been devoted to speeding up this process through subspace methods \citep{Harris2006, Harris2013, Barrett2014}, or fingerprint and similarity thresholding (FAST) \citep{Yoon2015}.
Recently, inter-station information has also been considered to improve phase picking efficiency and accuracy through inter-station coherence\citep{Delorey2017}, local similarity \citep{Li2018} and random sampling \citep{Zhu2017c}.

Facilitated by the parallel computation power of modern graphics processing units (GPUs), deep learning \citep{Goodfellow2016} took off for speech \citep{Hinton2012} and image recognition \citep{Krizhevsky2012} applications.
Most deep learning studies share the same fundamental network structure, such as the convolutional neural network (CNN), which further reduces the redundant model complexity of a neural network based on local conjunctions of features from the data (often found in images).
Unlike waveform similarity methods, CNNs trained on labeled datasets do not need a growing library of templates and seems to generalize well to waveforms not seen during training.
These recent developments have led to CNNs being applied to diverse seismic data sets\citep{Kong2018}, including volcanic events \citep{Luzon2017AGU}, induced seismicity \citep{Perol2018}, aftershocks \citep{Zhu2018}, as well as regular tectonic earthquakes recorded by regional seismic networks  \citep{Ross2018a,Ross2018b,ZhuBeroza2018}.
However, most of these works rely on a large volume of labeled training data which is only available in well-studied regions, such as California, US.

In this study, we accommodate the small seismic datasets by designing a specialized CNN network, named CNN-based Phase-Identification Classifier (\Acronym), for single-station multi-channel seismic waveforms.
The weights of the CNN are obtained via supervised training based on only thousands of human-labeled phase and non-phase samples used in a recent competition for detecting aftershocks of the 2008 \MW{7.9} Wenchuan earthquake in China \citep{Fang2017}.
The CNN learns a compact representation of seismograms in the form of a set of nonlinear local filters.
From the training process of discriminating seismic events from noise on large datasets, the weights of the local filters collectively capture the intrinsic features that most effectively represent seismograms for the given task of phase picking.
In the next sections, we show that \Acronym, trained on a much smaller labeled dataset, achieves comparable classification accuracy as reported in \cite{Ross2018b} and \cite{ZhuBeroza2018}.
\Acronym is further tested on a one-month continuous aftershock dataset for phase detection.
It achieves accurate detection of manually picked phases, precise arrival times of picked phases, as well as discovering many weak events not listed in the manual-picking catalog.

\section{Data}
\label{sec:data}

Unlike recent CNN studies that rely on an exceptionally rich training dataset of labeled samples \citep{ZhuBeroza2018, Ross2018b} to achieve good accuracy and robustness against noise, we design \Acronym and study its performance on a relatively small training set prior to applying it on a large volume of unlabeled data.
This is a typical scenario when analyzing the aftershock dataset of a major earthquake: strong aftershocks at a later time can be easily picked by existing algorithms or analysts; however, the real targets are the numerous number of aftershocks right after the mainshock that are missed by traditional methods \citep{Kagan2004,Peng2006}.
Prior to CNN training and processing, the only pre-processing applied to the seismogram is soft-clipping via a logistic function which is used to normalize the large dynamic range of the input waveforms.
As shown in \ref{sec:preprocessing}, such pre-processing contributes to \Acronym's stable convergence as well as higher accuracy.
Notably, no filtering is applied to the seismic waveforms in pre-processing.
\begin{figure}[t!]
    \centering
    \includegraphics[width=\linewidth]{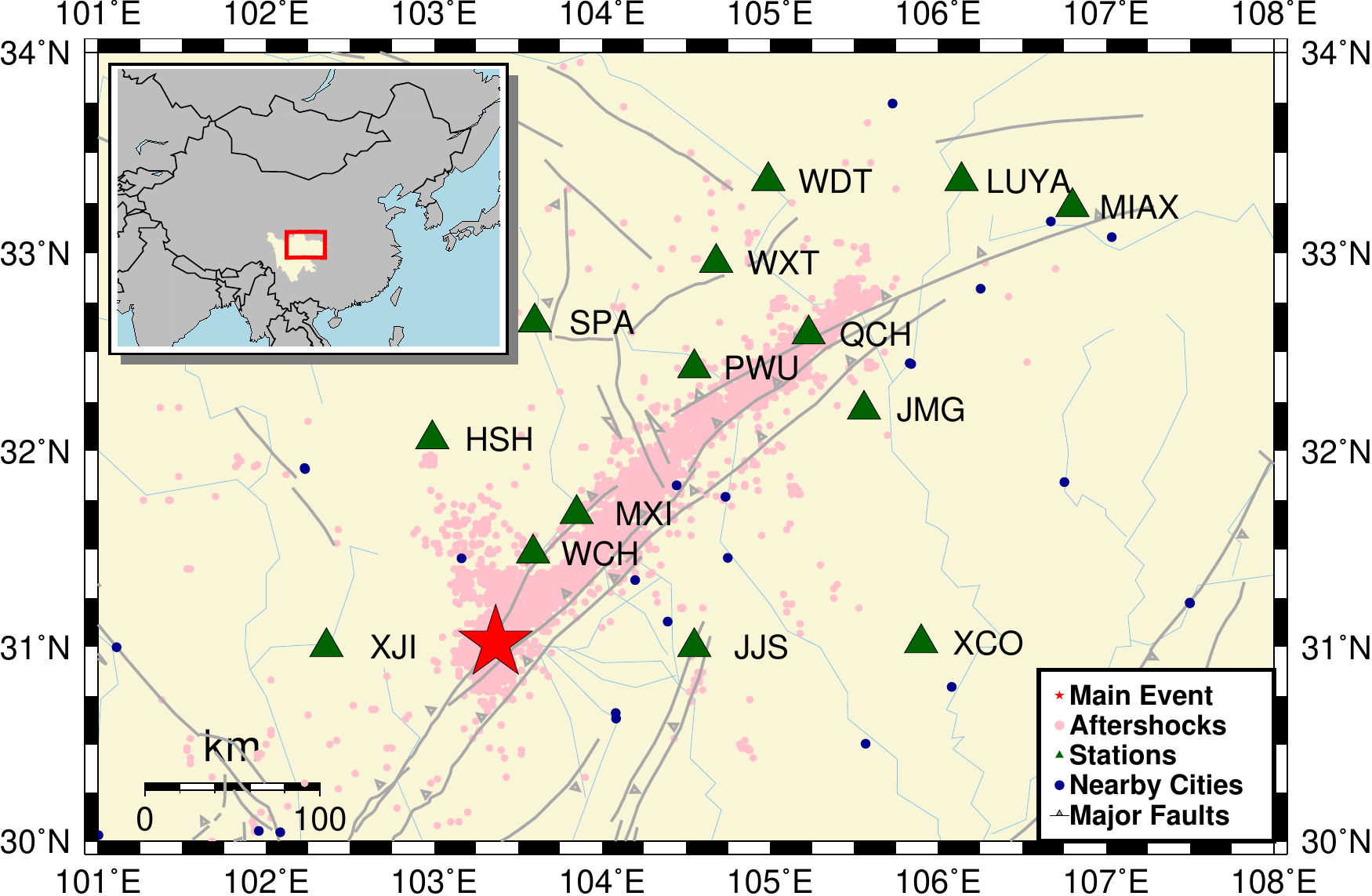}
    \caption{Map showing the study region in Sichuan, China along the aftershock zone of the 2008 \MW{7.9} Wenchuan earthquake (red star). The 9,361 manually picked aftershocks are marked as pink dots. The green triangles mark the 14 permanent stations that were used in this study. The gray and blue thin lines mark active faults and rivers in this region.}
    \label{fig:wenchuan-map}
\end{figure}
\begin{figure}[t!]
    \centering
    \begin{subfigure}{0.55\linewidth}
        \centering
        \includegraphics[width=\linewidth]{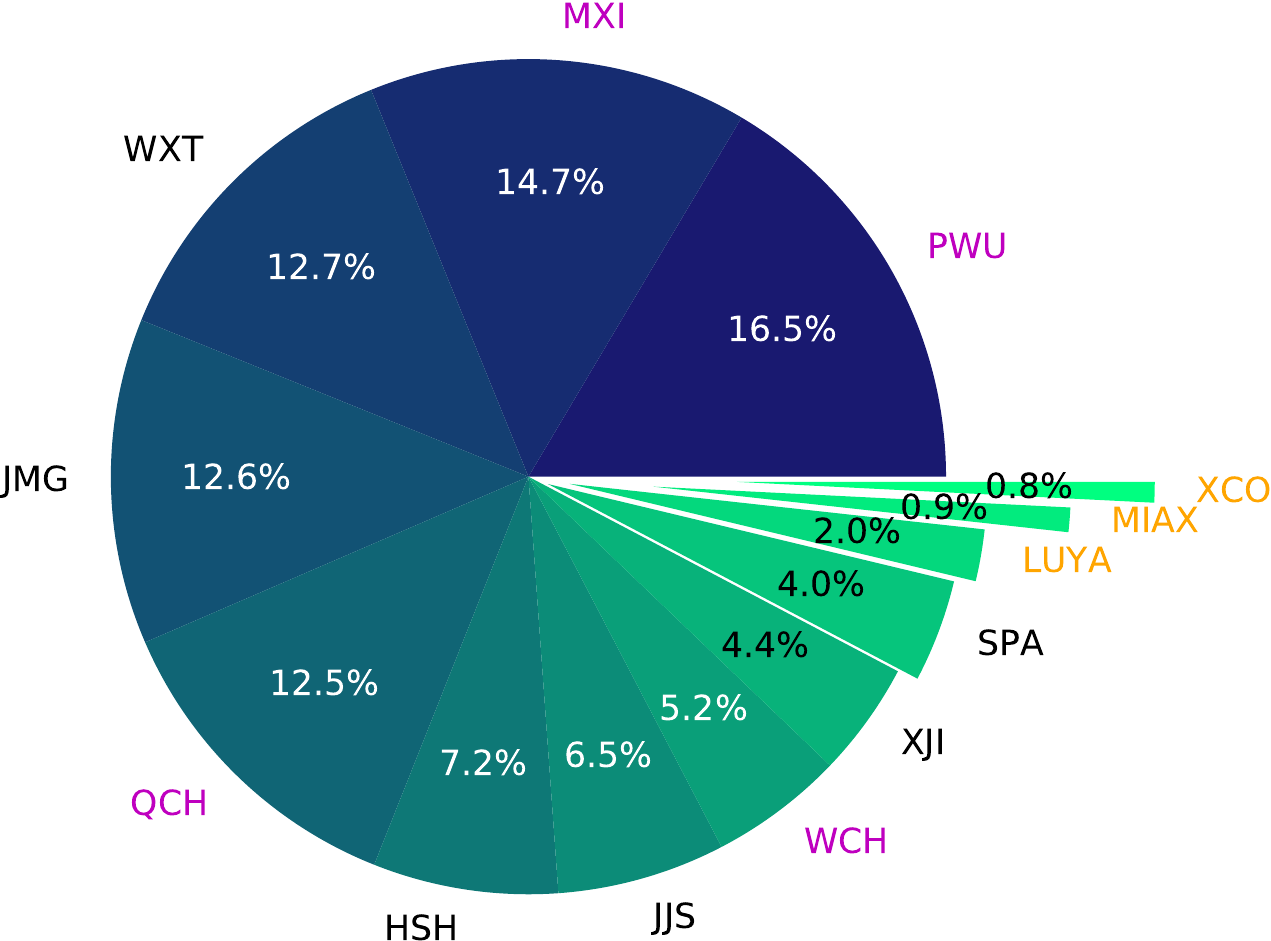}
        \caption{Event distribution over stations}
        \label{fig:event_distribution_stations}
    \end{subfigure}
    \hfill
    \begin{subfigure}{0.41\linewidth}
        \centering
        \includegraphics[width=\linewidth]{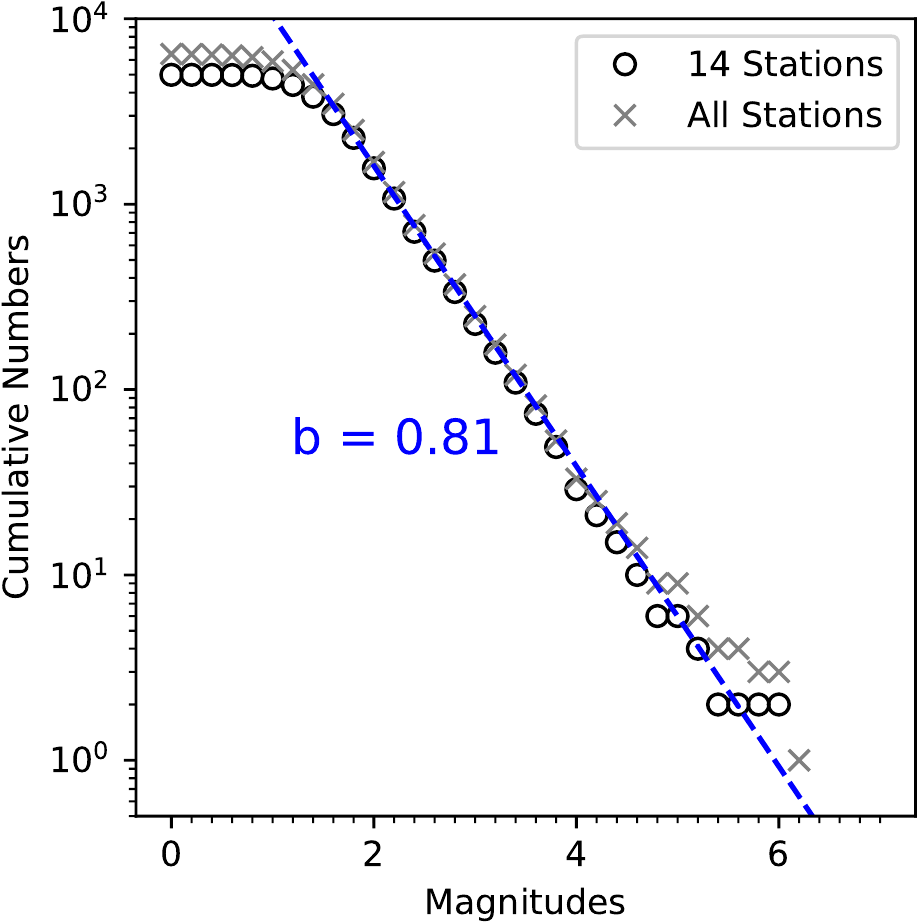}
        \caption{Event distribution over magnitudes}
        \label{fig:event_distribution_magnitudes}
    \end{subfigure}
    \begin{subfigure}{0.48\linewidth}
        \centering
        \includegraphics[width=\linewidth]{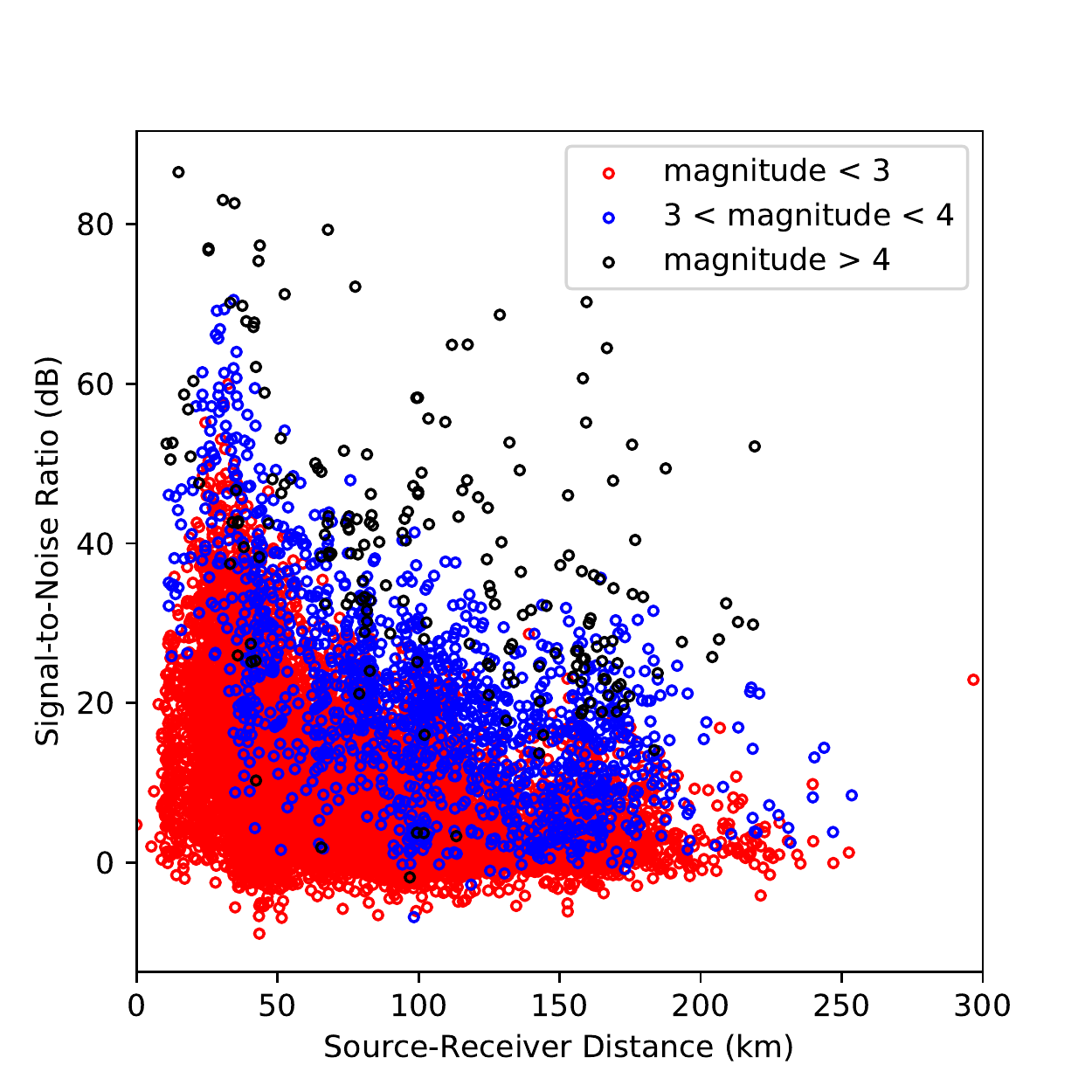}
        \caption{P phases}
        \label{fig:snr_p}
    \end{subfigure}
    \hfill
    \begin{subfigure}{0.48\linewidth}
        \centering
        \includegraphics[width=\linewidth]{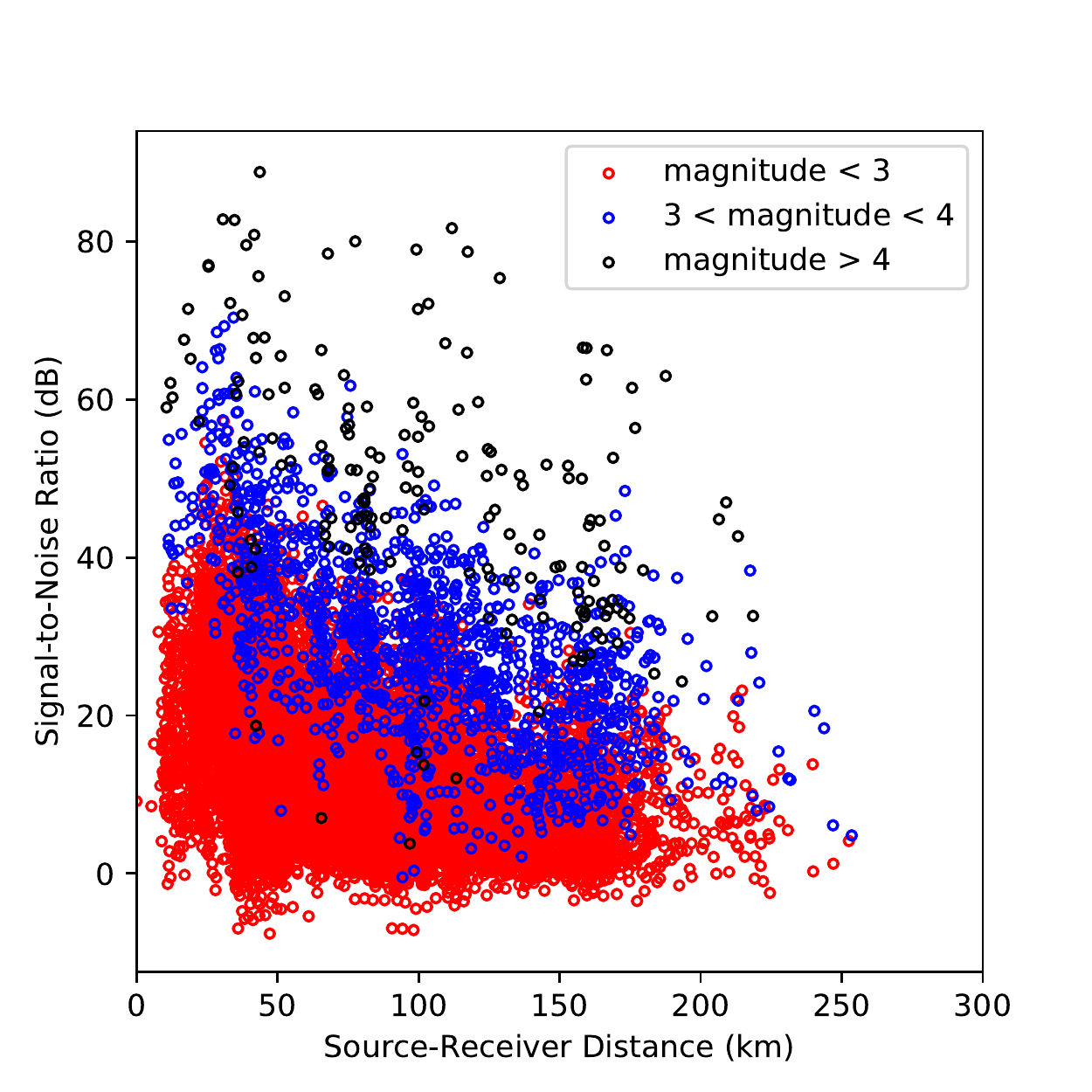}
        \caption{S phases}
        \label{fig:snr_s}
    \end{subfigure}
    \caption{Distribution of catalog events  in the Wenchuan aftershock dataset for (a) different stations and (b) different magnitudes. Stations on or close to the rupture zone are marked in purple while those far away are marked in gold. Signal-to-noise ratio of picked arrivals against event magnitudes and source-receiver distance for  (c) P phases and (d) S phases.}
    \label{fig:event_distribution}
\end{figure}

\paragraph{Study region}
We utilize the aftershock dataset of the 2008/05/12 \MW{7.9} Wenchuan earthquake that was made available during a recent competition for identifying seismic phases \citep{Fang2017}.
The mainshock occurred on the eastern margin of the Tibetan Plateau (Figure~\ref{fig:wenchuan-map}), and ruptured the central and northern section of the Longmenshan fault zone \citep{Xu2009,Feng2010,Hartzell2013}.
Numerous aftershocks occurred following the mainshock, but many of them were still missing in any published earthquake catalogs \citep{Yin2018}.
The aftershock dataset includes continuous data recorded for one month by 14 permanent stations in August 2008, which is three months after the Wenchuan mainshock.
Figure~\ref{fig:event_distribution_stations} shows the distribution of those phases among the 14 stations.
Stations near the aftershocks and the rupture zones (e.g., PWU, MXI, WXT, JMG, and QCH) had most of the picked phases, while distant stations (e.g., XCO, MIAX, LUYA, and SPA) have very few; and station \textsf{WDT} has no catalog phase arrivals.  % significantly less picked phases.

\paragraph{Catalogs}
The catalog we used contains 4,986 events with 30,146 phases manually picked on 14 permanent stations with arrivals of P (15,185) of S (14,961) phases.
Figure~\ref{fig:event_distribution_magnitudes} shows the catalog events distributed versus magnitude between \ML{0.3} to \ML{6.2}.
The signal-to-noise ratio (SNR) of each phase is computed as the ratio of signal powers between two 4-sec waveforms: one after each phase pick (signal) and one before its corresponding P arrival (noise).
Figures~\ref{fig:snr_p} and \ref{fig:snr_s} show the distribution of SNR of P and S phases against event magnitudes and source-receiver distance.
This catalog was used in a phase picking competition \citep{Zhu2017AGU} aiming to improve the detection and picking accuracy from the traditional methods.

\paragraph{Labeled dataset}
The \Acronym model is trained on a dataset of labeled seismic waveforms in 20-sec long windows.
\ref{sec:window_length} provides more details.
Adding noise-only windows, which are not included in the original labeled dataset, improves \Acronym's trained performance against noisy seismograms.
Here, we assume that quiet regions exist between 60\,s after an S-wave phase and 60\,s before a P-wave phase and generate 30,130 noise-only windows.
We note that because those noise windows were not verified manually, it is possible that they may include small aftershocks not listed in the catalog.
In the end, we obtain a dataset with 60,276 labeled windows, for which P-wave, S-wave, or noise labels have been assigned.

\paragraph{Continuous dataset}
Once \Acronym is trained on the labeled dataset, the phase detector and arrival picker are then tested on the entire one-month continuous waveforms starting on 08/01/2008 00:00:00 Beijing Time (or 07/31/2008 16:00:00 UTC).
Due to challenging acquisition conditions in the study area, there are some gaps in the continuous recording.
These are filled with zeros to keep the overall dataset consistent while avoiding artificial detections.

\section{Method}
\label{sec:method}

\begin{figure}[t!]
    \centering
    \includegraphics[width=\linewidth]{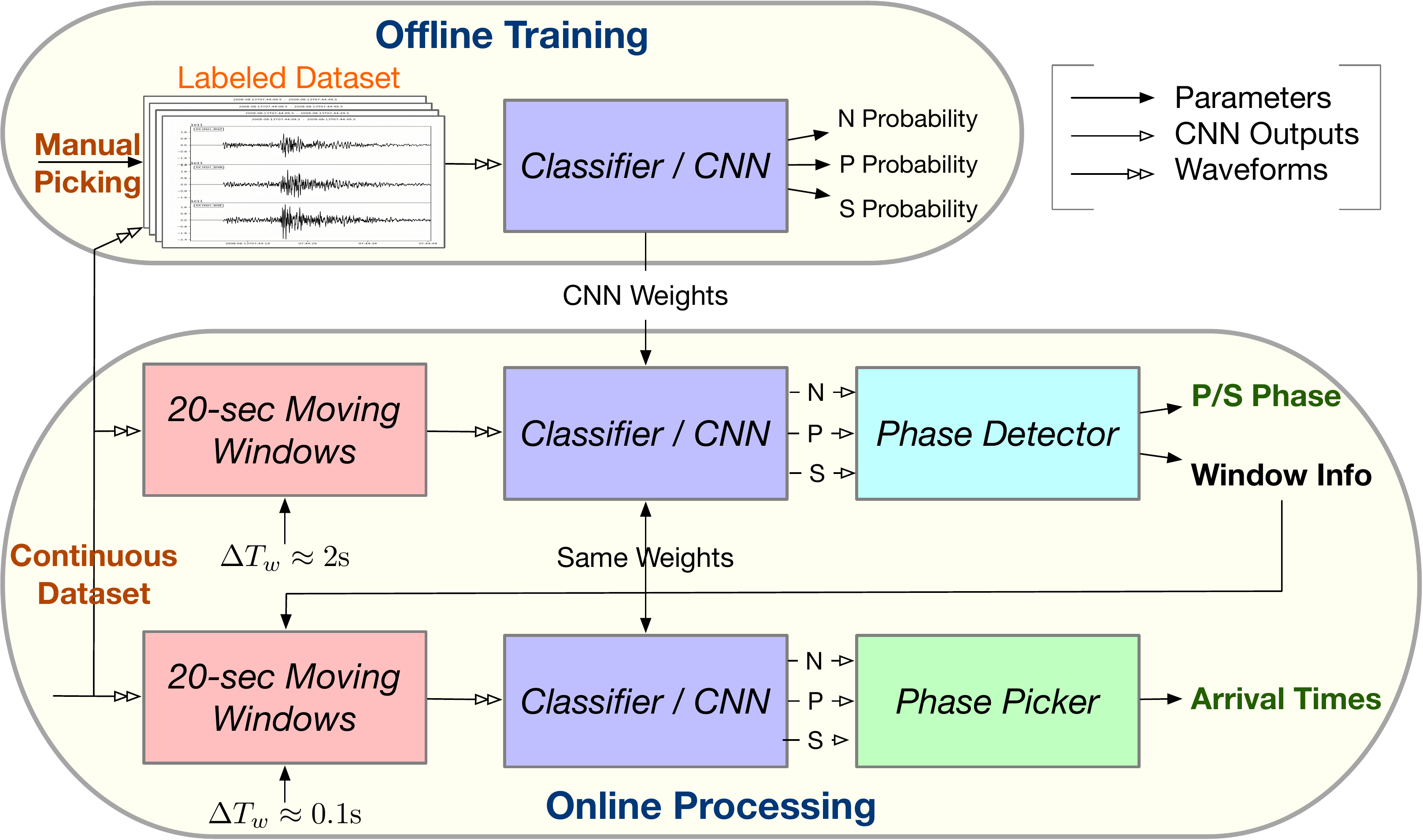}
    \caption{CNN-based Phase-Identification Classifier (\Acronym) flow chart. Inputs are three-component seismograms recorded at a single station, labeled in red. Outputs are P-wave, S-wave or noise window probabilities, and picked arrival times for P and S phases, shown in green. The 20-sec moving windows are overlapped with offsets controlled by $\Delta T_w$.}
    \label{fig:flow-chart}
\end{figure}

The task of finding a seismic phase and its arrival time is accomplished in two steps:
\begin{enumerate}
    \item \textit{Phase detection}: identify time windows where seismic phases exist;
    \item \textit{Phase picking}: determine the arrival times of the detected seismic phases within that time window.
\end{enumerate}
In this study, we adopt the processing pipeline summarized in Figure~\ref{fig:flow-chart}.
An off-line training process optimizes the parameters of the \textit{CNN-based classifier} iteratively over the labeled dataset.
The trained classifier is then used during on-line processing for both phase detection and picking.
The \textit{Phase detector} employs moving windows with 90\% overlap ($\Delta T_w = 2$\,s offset) and casts seismic phase detection as a classification problem \citep{Zhu2017b} of P-wave, S-wave, or noise-only labels.
The detected windows are then inputted to the same classifier to generate characteristic functions (CFs) on a finely sampled grid, e.g., $\Delta T_w = 0.1$\,s offset.
The \textit{phase picker} estimates the arrival times based on the peaks of smoothed CFs.
Multiple window offsets, $\Delta T_w$, were tested in a grid search manner.
In general, a smaller $\Delta T_w$ gives better picking accuracy; however, the computation cost is also inversely proportional to $\Delta T_w$.

\begin{figure}[t!]
    \centering
    \includegraphics[width=\linewidth]{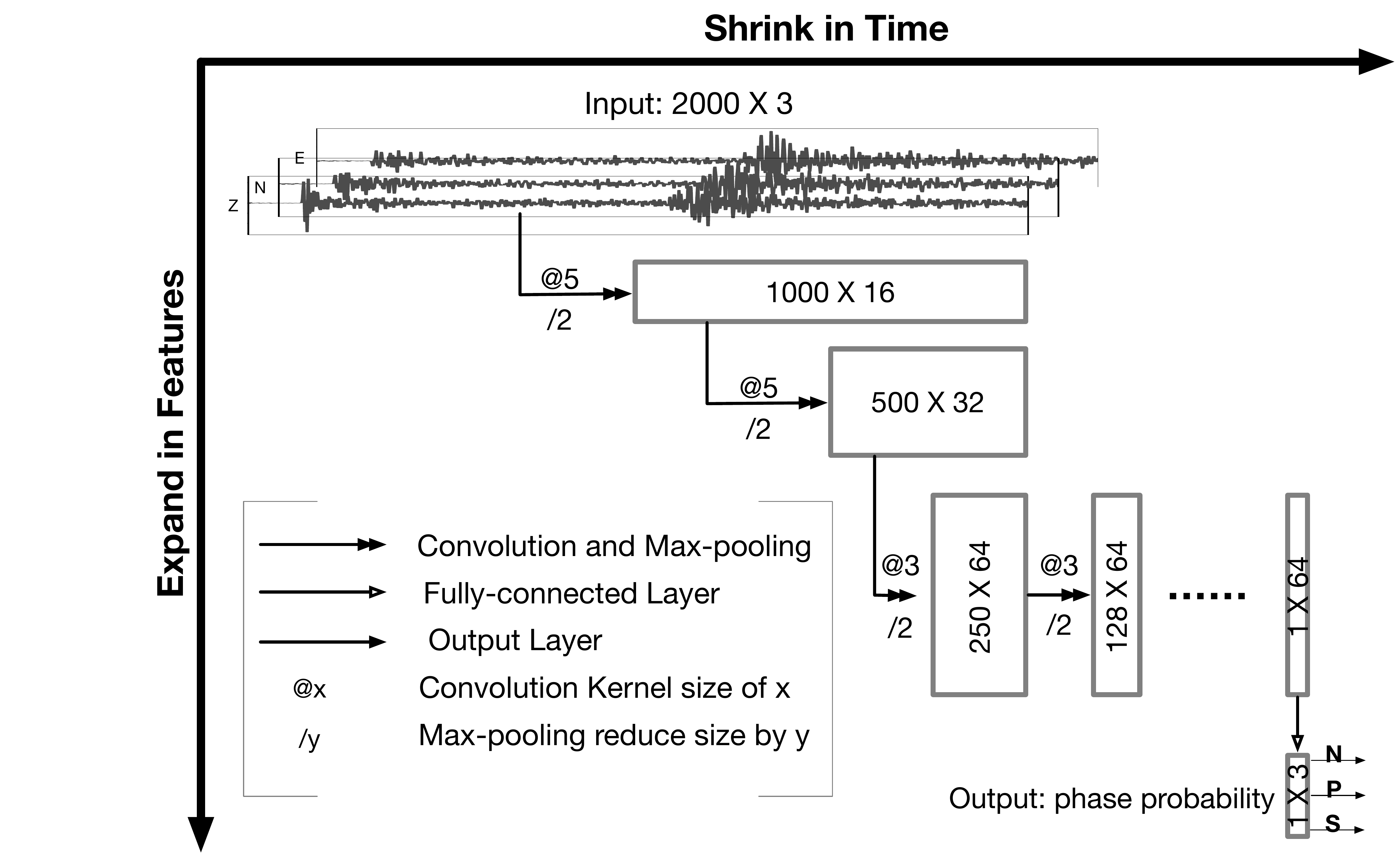}
    \caption{A diagram showing the CNN network structure. Each input is a 3-C seismogram (20-sec window) which shrinks in time but expands in the feature dimension as it passes through 11 convolutional layers for feature extraction. The final layer is fully connected with 3 outputs that give the probabilities of a window being noise, P, and S phases.}
    \label{fig:cnn}
\end{figure}
%

%% Our design
\subsection{CNN-based Classifier}
\label{sec:method-cnn}

The classifier in Figure~\ref{fig:flow-chart} operates on inputs that are 3-C seismograms in 20-s windows, sampled at 100\,Hz.
Its outputs are probabilities of each window containing a P/S phase arrival at 5\,s, or only noise.
The CNN classifier contains 11 convolutional layers along with one fully-connected layer (Figure~\ref{fig:cnn})
It is trained by processing many labeled windows known to contain P or S phases, or noise only.

A \textit{Softmax} function is used to normalize the probabilities in the output layer:
\begin{equation}
    q_i(x) = {e^{z_i(x)}}/(e^{z_0(x)}+e^{z_1(x)}+e^{z_2(x)})
    \label{eq:softmax}
\end{equation}
where $i=0,1,2$ represents noise, P, and S classes, and $z_i(x)$ is the unnormalized output of the last fully-connected (FC) layer for the $i^\text{th}$ class.
A loss function is needed when optimizing the CNN weights during the training  process, so we use the cross-entropy between a true probability distribution $p$ and the estimated distribution $q$ which is defined as
\begin{equation}
    H(p, q) = -\sum_x p(x) \log q(x)
    \label{eq:xentropy}
\end{equation}
Hence, the \textit{Softmax} classifier minimizes the cross-entropy between the estimated class probabilities ($q$ defined in \eqref{eq:softmax}) and the “true” distribution, which is the distribution where all probability mass is on the correct class, e.g., $p = (0, 1, 0)$ for a labeled P phase window.
Between each layer, a rectified linear unit (ReLU) activation function \citep{Nair2010} introduces nonlinearity into the model.
The data size is reduced at each layer using max-pooling \citep{Zhou1988}.
% Optimization of the training problem is performed using Adam \citep{Kingma2014}.
% Batch normalization \citep{Ioffe2015} is also integrated to further stabilize the training process.

To accommodate small to medium training set sizes, the proposed CNN uses only one convolution layer between each max-pooling layer.
This results in 107,248 parameters in the CNN for a 20-sec window length.
The number of parameters can be reduced if a shorter window length is chosen instead.
Since each layer down-samples the input data by a factor of two, the model can adjust to a different window length by adding or removing layers.
Finally, the number of FC layers used here is fewer than commonly seen in CNNs.
We experimented with different numbers of FC layers (one, two, and three) but found no discernible difference in the classifier accuracy.
Thus, we chose the structure with fewest FC layers for the sake of simplicity.

\begin{figure}[t!]
    \captionsetup[subfigure]{skip=-4mm,justification=justified,margin=6mm,singlelinecheck=false}
    \centering
    \begin{subfigure}{\columnwidth}
        \centering
        \caption{}
        \label{fig:cpic_demo_waveform}
        \includegraphics[width=\columnwidth]{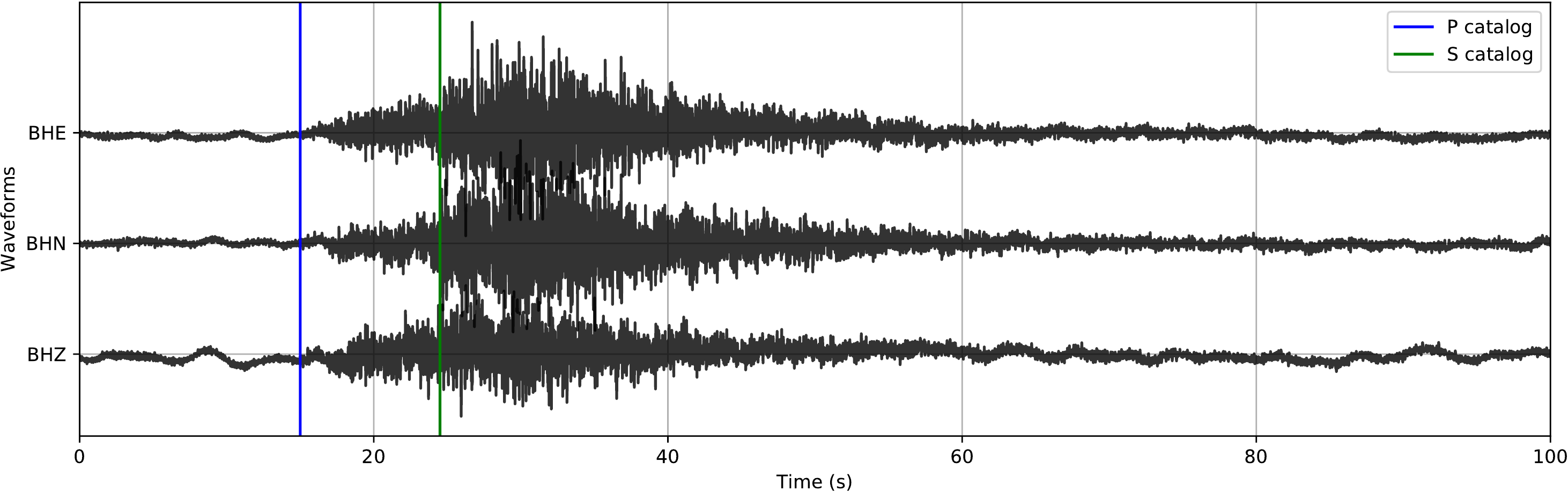}
    \end{subfigure}
    
    \begin{subfigure}{\columnwidth}
        \centering
        \caption{}
        \label{fig:cpic_demo_detect}
        \includegraphics[width=0.99\columnwidth]{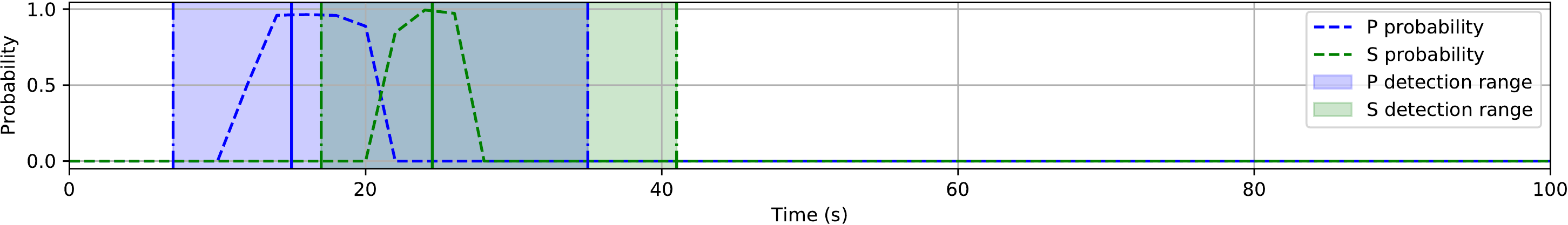}
    \end{subfigure}
    
    \begin{subfigure}{0.48\columnwidth}
        \centering
        \caption{}
        \label{fig:cpic_demo_pick_p}
        \includegraphics[width=\columnwidth]{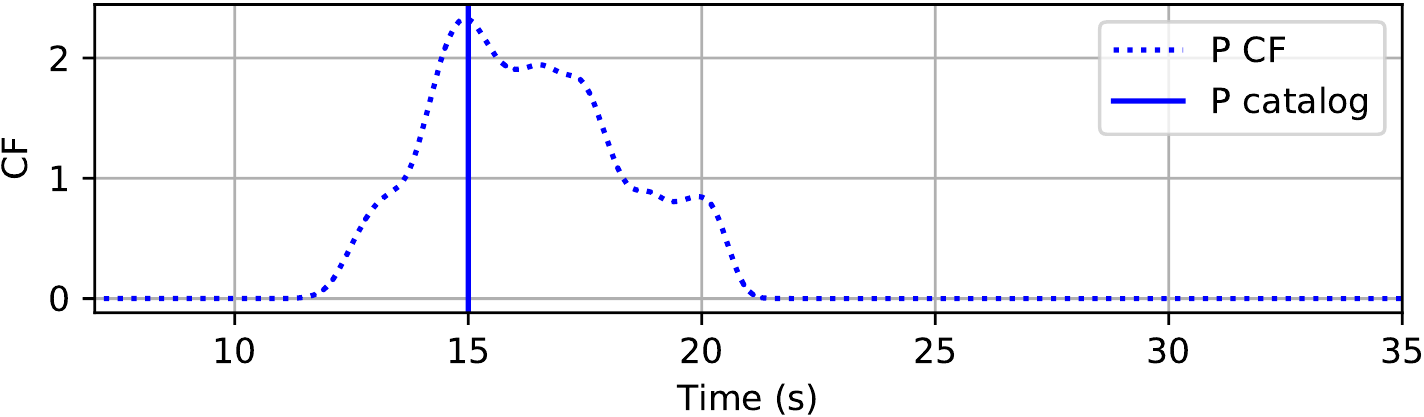}
    \end{subfigure}
    \hfill
    \begin{subfigure}{0.48\columnwidth}
        \centering
        \caption{}
        \label{fig:cpic_demo_pick_s}
        \includegraphics[width=\columnwidth]{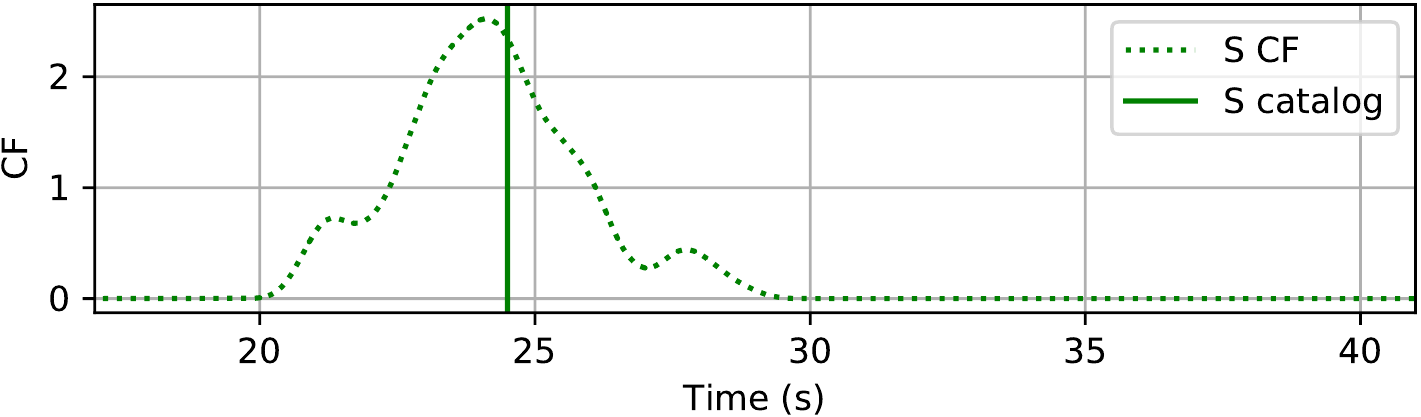}
    \end{subfigure}
    \caption{\Acronym work flow: (a) Three-component waveforms (catalog P and S arrivals marked) are taken as input; (b) probabilities of both P and S phases are calculated every 2\,s from which the P and S detection ranges (shaded) are selected, starting 5\,s before the first nonzero probability sample, and ending and 15\,s after the last. (c, d) Arrival times are picked on characteristic functions (CFs) calculated every 0.1\,s within each detection range in (b).}
    \label{fig:cpic_demo}
\end{figure}
\begin{figure}[t!]
    \centering
    \includegraphics[width=\columnwidth]{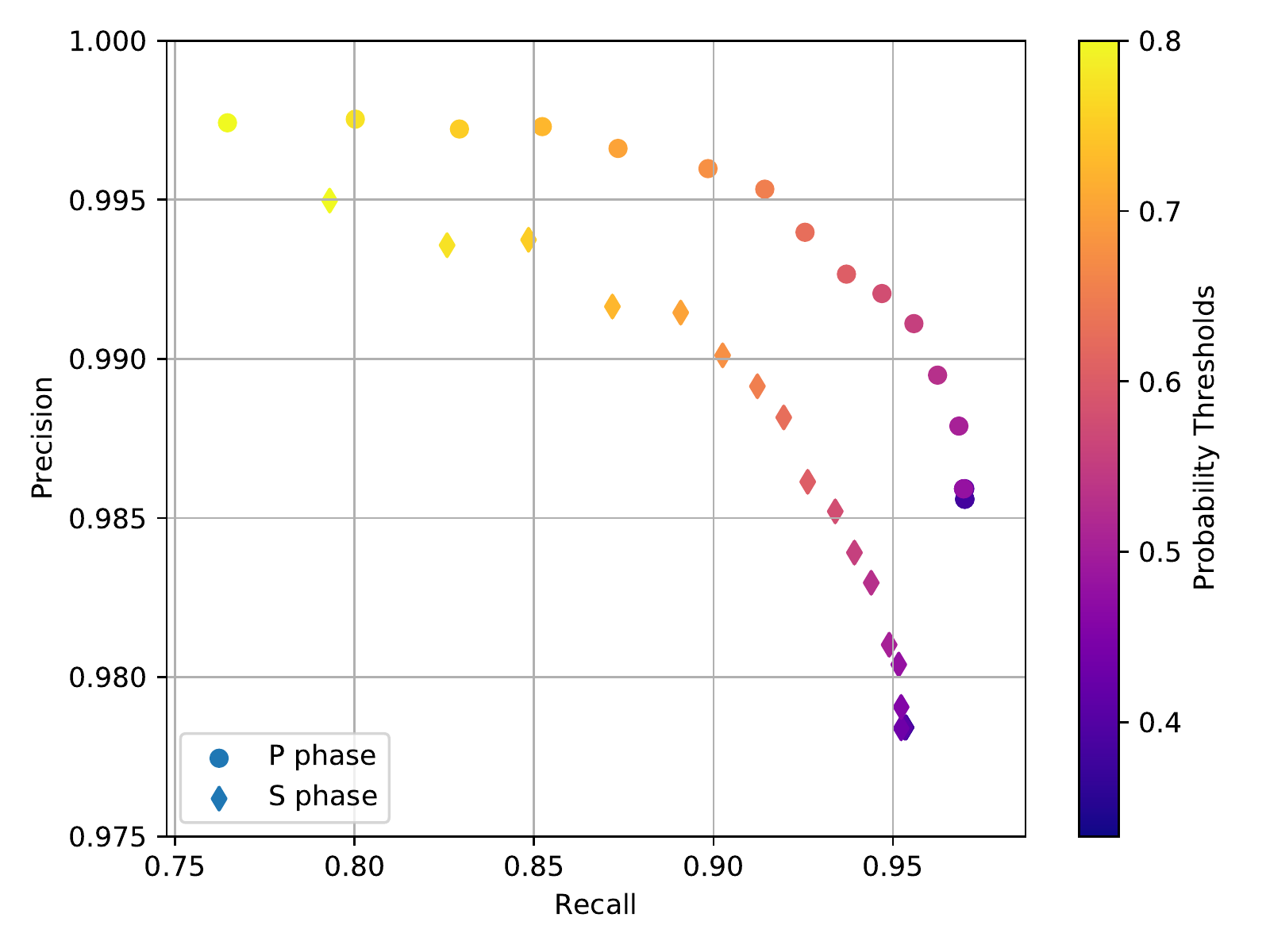}
    \caption{Precision-recall curve for P and S phase detection under different probability thresholds. The top left is the high-precision-low-recall region and the bottom right is the low-precision-high-recall region. A threshold of 0.5 gives the highest precision with recall larger than 0.95. Only P or S phases with a probability higher than both the noise and the threshold are valid detections. This results in the effective minimum threshold at 0.33 for this tri-class classifier.}
    \label{fig:precision_recall}
\end{figure}
\subsection{Phase Detector}
\label{sec:method-phasedetector}

The phase detector in Figure~\ref{fig:flow-chart} for continuous processing works on the CNN classifier outputs from moving windows that are coarsely sampled.
The three outputs from the CNN classifier are converted to probabilities of noise, P phase, and S phase at each window position by \eqref{eq:softmax}.
A peak probability above $0.5$ is sufficient for detecting a P-phase or S-phase window.
Every positive detection provides a candidate 20-sec window that may contain P or S phases.
Overlapping windows with the same phase label are merged into one longer window before passing to the phase picker.
A detection example of a typical 100-sec waveform is provided in Figure~\ref{fig:cpic_demo_detect}.

The threshold $0.5$ for event detection is chosen from the precision-recall tradeoff curve shown in Figure~\ref{fig:precision_recall} because it gives the highest precision with a recall larger than 0.95.
Notice that one can remove the constraint that a detected phase needs to have a probability higher than the noise class when weak events are sought in a low-SNR scenario.
However, this practice, which increases the false alarm rate and results in a lower precision, is not recommended.
This low-precision-high-recall region is not shown in Figure~\ref{fig:precision_recall}), but it would extend the curve further to the right.
Note that the confusion matrix shown in Table~\ref{tab:confusion-matrix} reflects the best amount of data points for P and S phases in this plot.

\subsection{Phase Picker}
\label{sec:method-phasepicker}

The phase picker in Figure~\ref{fig:flow-chart} recomputes the CNN classifier outputs over the detected windows with a smaller offset to obtain the resolution needed for accurate time picking.
Since the window of P and S phases starts 5\,s prior to the picked arrival time, the probabilities output from the CNN classifier also reflect the likelihood of phase arrivals at 5\,s of the given window.
Thus, the probability of each phase (the arrival time at 5\,s of the corresponding window) should reach a local peak at the true arrival time.
Instead of using the probabilities of each phase directly, the \textit{phase picker} relies on characteristic functions (CFs) computed as the smoothed log ratio between probabilities of each phase against the noise class.
Using a ratio between phase and noise probabilities makes the constructed CFs adaptive to corresponding noise levels.
This helps to eliminate false picks caused by background noise.
Picking examples of P and S phases on the detected windows from Figure~\ref{fig:cpic_demo_detect} are given in Figure~\ref{fig:cpic_demo_pick_p} and \ref{fig:cpic_demo_pick_s}, respectively.
Comparing to the probabilities in Figure~\ref{fig:cpic_demo_detect}, CFs emphasize the arrival times of P and S phases and suppress the significance of their coda waves.

However, it is possible that multiple picks are present in one single detection window.
\Acronym does not force a single pick in one window; instead, it assigns a confidence level to each pick.
This confidence is measured by the peaks' relative \emph{prominence}, which is defined as the vertical distance between the peak and its lowest contour line \citep{Helman2005finest}.
This measure makes the picking process parameter-free; however, one can specify a minimum confidence level (e.g., $1/(n+1)$ where $n$ is the number of picks) for a multiple-pick scenario.
For example, three picks with confidences level as $(0.4, 0.45, 0.15)$.
A $0.25$ threshold of confidence rejects the pick with $0.15$ prominence while keeping the first two picks.
Notice that setting a $0.5$ confidence threshold effectively forces a single pick in a detection window.

\section{Performance Evaluation}
\paragraph{CNN Classifier}
%As mentioned before, the CNN classifier categorizes 20-sec time windows into noise, P phase or S phase classes.
We can evaluate a CNN classifier by processing labeled testing data where the true output is known.
The \emph{accuracy} defined below is a simple measure of a classifier's performance:
\begin{equation}
    \text{accuracy} = \frac{\text{number of correctly labeled samples}}{\text{number of all testing samples}}
    \label{eq:accuracy}
\end{equation}
Noise labels are not treated differently from phase labels, so classifying a noise window correctly has the same weight as confirming a phase window. %as well as deny a noise window.

\paragraph{Phase Detector}
The detector can be viewed as a three-class classifier that decides whether a given time window contains a seismic phase (P or S), or only noise.
%A label of specific seismic phase is also given as a by-product of using a multi-class classifier as shown in Figure~\ref{fig:flow-chart}.
To evaluate the detector's effectiveness, we use a confusion matrix as in Table~\ref{tab:confusion-matrix}, where the labeled windows of each class (per row) are sorted into the number of each detected type (per column).
Subscripts denote the detected class, e.g., $P_s$ is the number of windows with P-phase labels but detected as S-phase.
The sum of all nine counts equals the total number of labeled windows in the given catalog.
%, which is typically very large
% % when applying a phase detector on
%for a continuous dataset.
\begin{table}
    \caption{Definition of confusion matrix for evaluating phase detector}
    \label{tab:confusion-matrix}
    \centering
    \begin{tabular}{l|l|c|c|c|c}
        \multicolumn{2}{c}{}&\multicolumn{3}{c}{Detector}&\\
        \cline{3-5}
        \multicolumn{2}{c|}{}&Noise&P-wave&S-wave&\multicolumn{1}{c}{Total}\\
        \cline{2-5}
        \multirow{3}{*}{\rotatebox{90}{Catalog}}& Noise & $N_n$ & $N_p$ & $N_s$ & $N_n + N_p + N_s$\\
        \cline{2-5}
        & P-wave & $P_n$ & $P_p$ & $P_s$ & $P_n + P_p + P_s$\\
        \cline{2-5}
        & S-wave &$S_n$ & $S_p$ & $S_s$ & $S_n + S_p + S_s$\\
        \cline{2-5}
        \multicolumn{1}{c}{} & \multicolumn{1}{c}{Total} & \multicolumn{1}{c}{$N_n + P_n + S_n$} & \multicolumn{1}{c}{$N_p + P_p + S_p$}  & \multicolumn{1}{c}{$N_s + P_s + S_s$} & \multicolumn{1}{c}{$ALL$}\\
    \end{tabular}
\end{table}
%Since most windows in the continuous dataset are noise windows, $T_n$ is very large, but not important when evaluating performance.
To avoid the effect of an imbalanced dataset dominated by noise windows (large $N_n$), we can use $precision$ and $recall$ (a.k.a. sensitivity) for each class to measure the performance, which ignores  $N_n$.
% of the phase detector and ignore $N_n$.
%For example, $precision$ and $recall$
These are defined for the P-wave class as:
\begin{equation}
    \begin{aligned}
        &precision: &\mathcal{P}_p = \frac{P_p}{N_p + P_p + S_p}\\
        &recall: &\mathcal{R}_p = \frac{P_p}{P_n + P_p + P_s}\\
    \end{aligned}
\label{eq:pr}
\end{equation}
$\mathcal{P}_n, \mathcal{P}_s, \mathcal{R}_n$ and $\mathcal{R}_s$ can be defined similarly.
Notice that both $precision$ and $recall$ are independent of $N_n$.
Ideally, both $\mathcal{P}$ and $\mathcal{R}$ for each class would be close to 1,
However, the labeled aftershock dataset catalog we have is incomplete\,--\,it tends to include only the strong and obvious phases while omitting weak events.
Thus, higher $N_p$ and $N_s$ counts are expected which lowers $\mathcal{P}_p$ and $\mathcal{P}_s$, although some of these $N_p$ and $N_s$ detections are likely weak phases not listed in the catalog.
On the other hand, $\mathcal{R}_p$ and $\mathcal{R}_s$ should be high if very few manually labeled strong phases are missed.
Notice that the accuracy defined in \eqref{eq:accuracy} measures the ratio between the sum of diagonal terms over all terms in the confusion matrix:
\begin{equation*}
    accuracy = \frac{N_n + P_p + S_s}{ALL}
\end{equation*}
Similarly, to avoid a dominant $N_n$ count biasing the accuracy, the F-1 score is computed from $precision$ and $recall$ (their harmonic mean) for each class:
\begin{equation}
    \text{F-1}= \left(\frac{precision^{-1} + recall^{-1}}{2}\right)^{-1}
\end{equation}

\paragraph{Phase Picker}
The phase picking process estimates the arrival time for each detected seismic phase.
We measure our phase picker's error as
%the difference between our arrival times and the catalog ones:
% by subtracting catalog arrival times from our picks:
\begin{equation}
    E_\text{pick} = T_\text{pick} - T_\text{cat}
    \label{eq:error}
\end{equation}
where $T_\text{pick}$ is the arrival time from \Acronym  and $T_\text{cat}$ is the manually picked phase arrival time.
Then the systematic bias and variance of our phase picker estimator are measured by taking the mean and standard deviation of $E_\text{pick}$ over all catalog phases.
%A close-to-zero bias is expected for good manual picking cases.
%Also, we expect a reasonably low variance even though the manual pick may contain some human error.
We expect a close-to-zero bias and reasonably low variance even though the catalog pick itself may contain some human error.
Note that the catalog phase arrival time is rounded to the tenth decimal point (0.1).

\section{Results}
\label{sec:results}

\subsection{Training and testing of the CNN classifier}
\label{sec:results-cnn}

\begin{figure}[t!]
    \centering
    \includegraphics[width=\columnwidth]{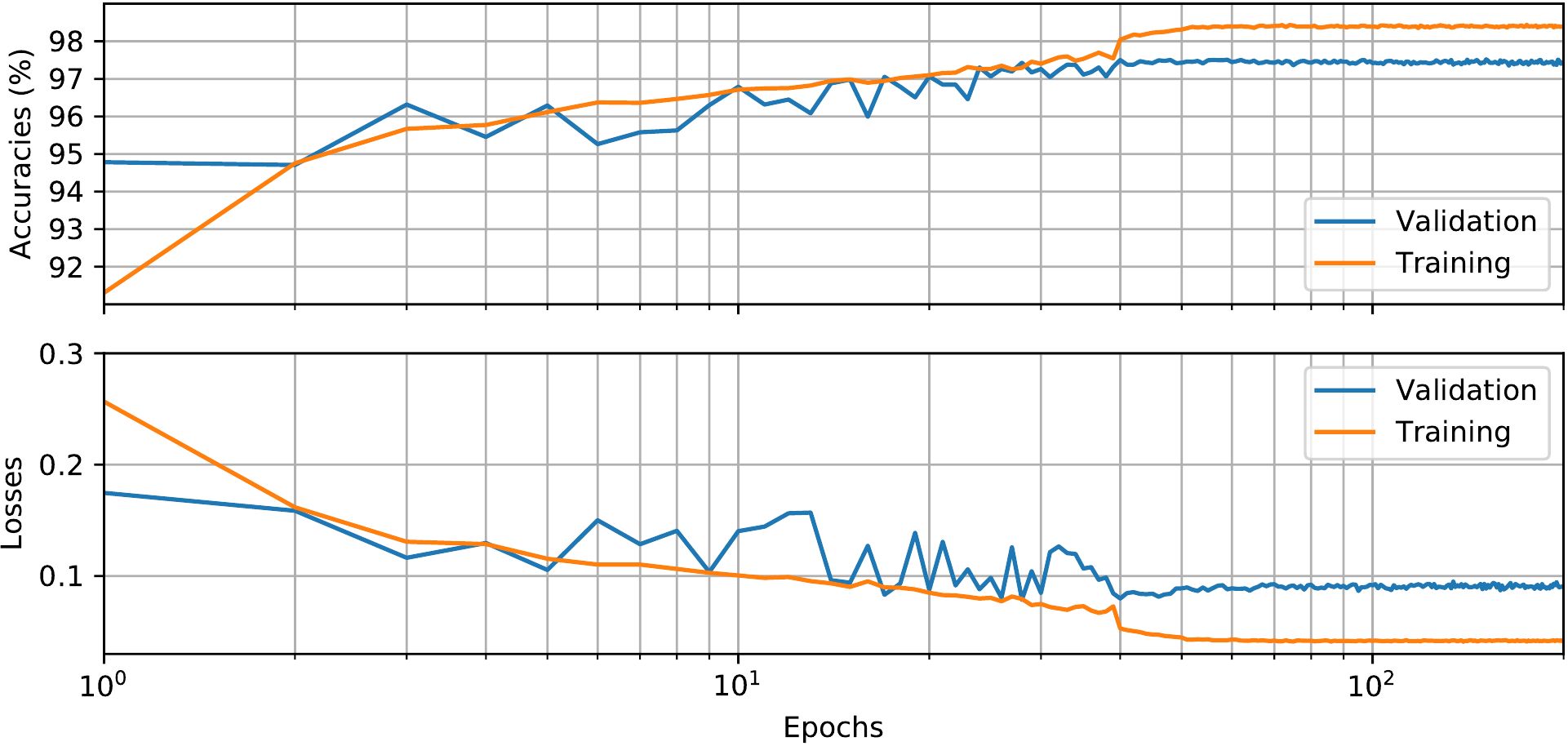}\llap{
        \parbox[b]{11cm}{\soulwhite{(a)}\\\rule{0ex}{2.5cm}(b)\\\rule{0ex}{2.5cm}}
    }
    \caption{Training performance: (a) classifier accuracy and (b) loss function against number of epochs on training and validation datasets during the CNN training process.}
    \label{fig:evaluate_train}
\end{figure}

\begin{figure}[t!]
    \captionsetup[subfigure]{skip=-0.5mm}
    \centering
    \begin{subfigure}{0.32\linewidth}
        \includegraphics[width=\columnwidth]{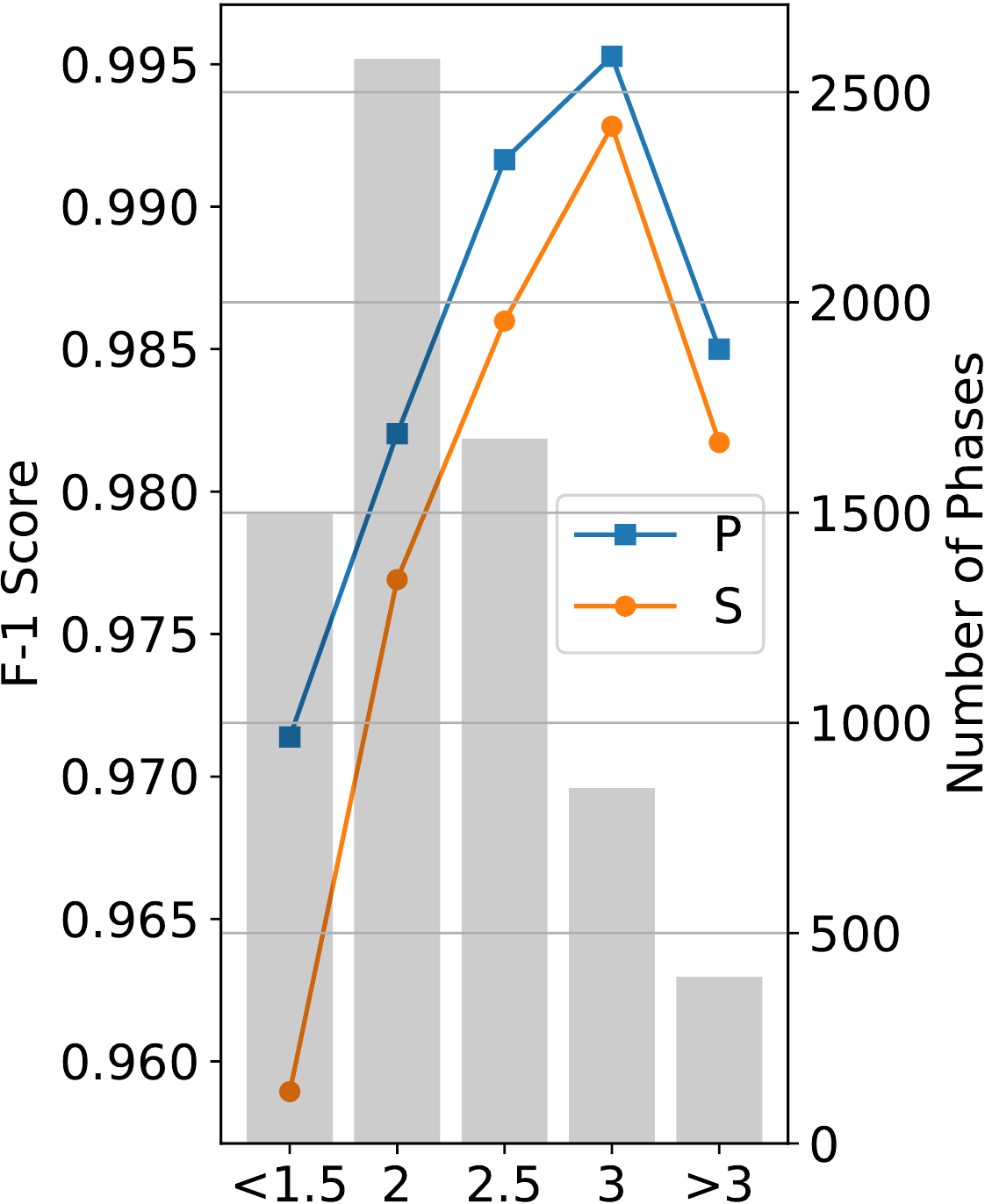}
        \caption{Magnitude}
        \label{fig:f1_vs_mag}
    \end{subfigure}
    \hfill
    \begin{subfigure}{0.32\linewidth}
        \includegraphics[width=\columnwidth]{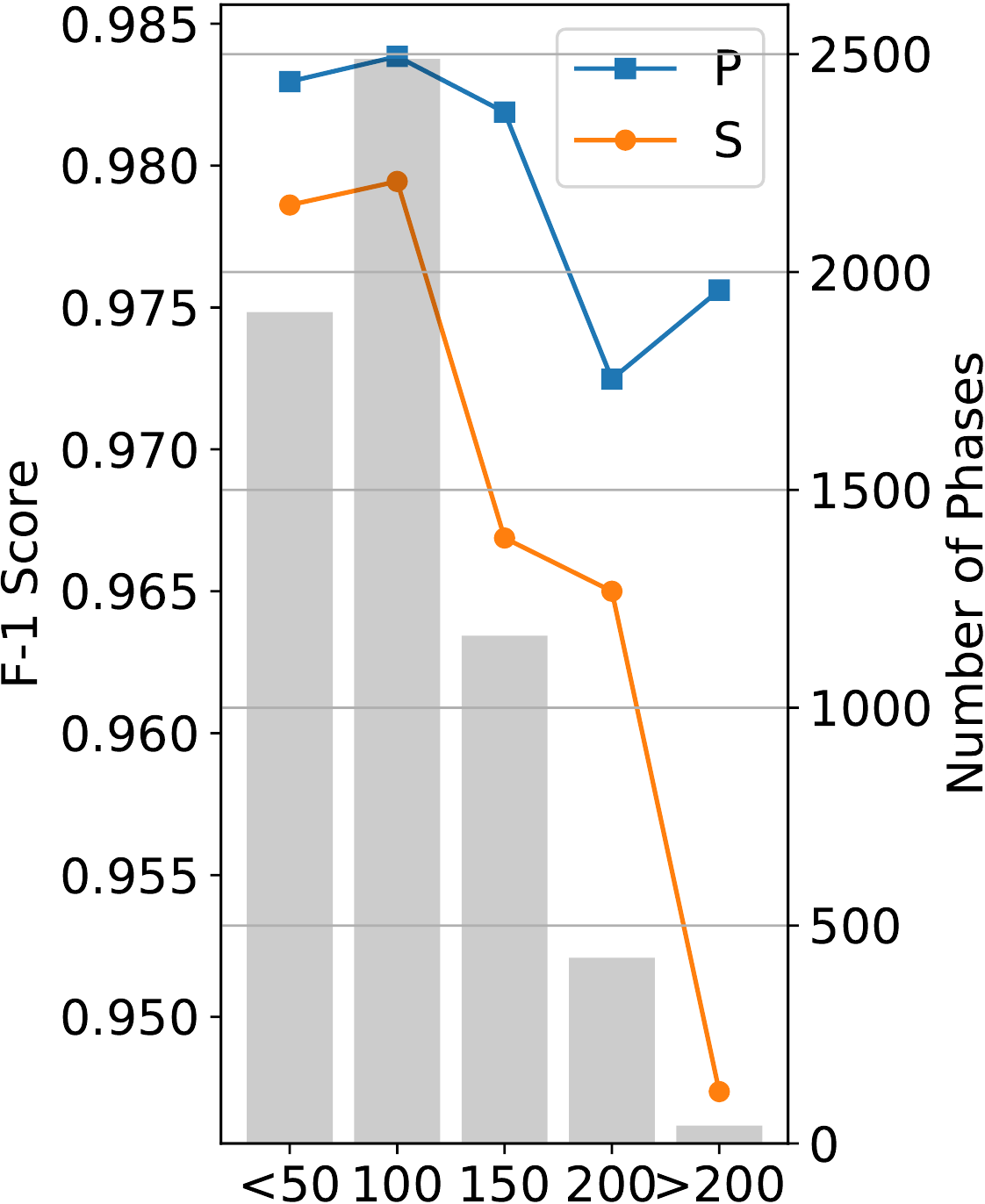}
        \caption{Distance}
        \label{fig:f1_vs_dist}
    \end{subfigure}
    \hfill
    \begin{subfigure}{0.32\linewidth}
        \includegraphics[width=\columnwidth]{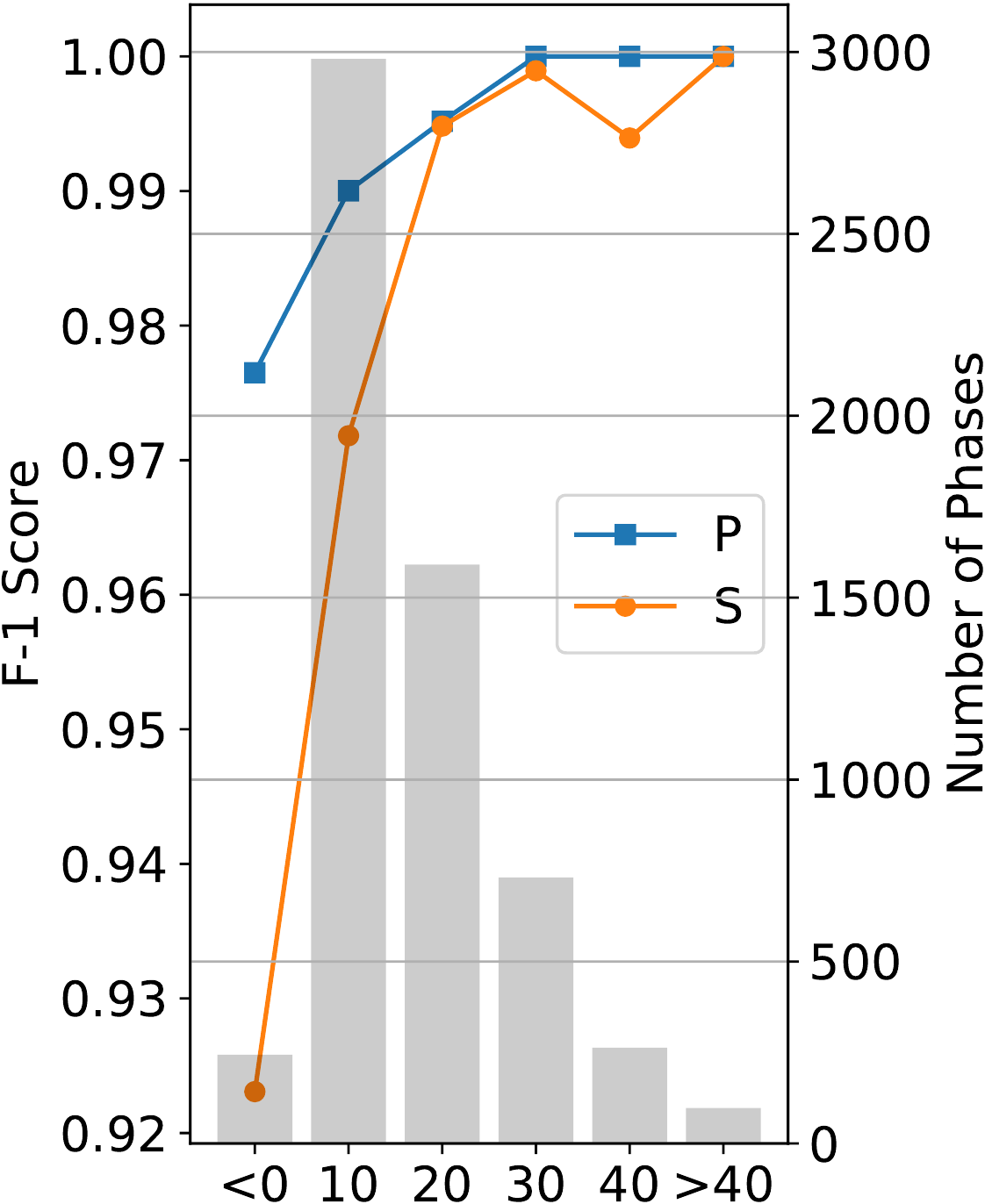}
        \caption{SNR}
        \label{fig:f1_vs_snr}
    \end{subfigure}
    \caption{F1 scores (right axes) of the trained classifier versus (a) magnitude, (b) distance, and (c) SNR. P (blue) and S (orange) phases are plotted separately. The number of testing samples in each small bin (left axes) is shown by the bars in the background. }
    \label{fig:performance_analysis}
\end{figure}

% Training
To systematically verify the accuracy and stability of the proposed CNNs, the available 60,000 labeled windows are split into a training subset and a testing subset.
The split is done chronologically to emulate a real-world scenario: training on historical phases (80\%) and testing on future ones (20\%).
The training process involves minimization of the loss function \eqref{eq:xentropy} with iterative updating based on the gradient.
After the CNN training process sees every sample in the entire training dataset once, we have finished one \emph{epoch} of training.
At the end of each epoch, we generate a testing result to score the CNN classifier accuracy and thus track the progress of its training.
Multiple epochs are needed to fully train the CNN weights into a stable state.
\begin{table}
    \caption{Confusion matrix for phase classification on the validation dataset which is the latest 20\% of the labeled phases.}
    \label{tab:confusion_matrix}
    \centering
    \begin{tabular}{l|l|c|c|c|c}
        \multicolumn{3}{c}{}&\multicolumn{2}{c}{Detector}&\\
        \cline{3-5}
        \multicolumn{2}{c|}{}&Noise&P-wave&S-wave&Total\\
        \cline{2-5}
        \multirow{3}{*}{\rotatebox{90}{Catalog}}& Noise & $5,946$ & $97$ & $113$ & $6,156$\\
        \cline{2-5}
        & P-wave &$22$ & $2,930$ & $10$ & $2,962$\\
        \cline{2-5}
        & S-wave &$59$ & $6$ & $2,873$ & $2,938$\\
        \cline{2-5}
        \multicolumn{1}{c}{} & \multicolumn{1}{c}{Total} & \multicolumn{1}{c}{$6,027$} & \multicolumn{1}{c}{$3,033$} & \multicolumn{1}{c}{$2,996$} & \multicolumn{1}{c}{$12,056$}\\
    \end{tabular}
\end{table}
\begin{table}
    \caption{Precision, recall, and F-1 score for the three classification categories.}
    \label{tab:performance_scores}
    \centering
    \begin{tabular}{lccc}
        \toprule
        Categories & Precision & Recall & F-1 Score\\
        \midrule
        Noise & $0.987$& $0.966$ & $0.976$\\
        P-wave & $0.966$ & $0.989$ & $0.9787$\\
        S-wave & $0.959$ & $0.978$ & $0.968$\\
        \bottomrule
    \end{tabular}
\end{table}
\paragraph{Reliable classifier}
As demonstrated in Figure~\ref{fig:evaluate_train}, the training process of the proposed CNN converges after 40 epochs; no over-fitting is observed even after 200 epochs.
The overall validation accuracy of this experiment reaches 97.5\%, using the diagonal entries of detailed confusion matrix shown in Table~\ref{tab:confusion_matrix}.
Precision, recall, and F-1 scores are given in Table \ref{tab:performance_scores}.
To further understand characteristics of the trained CNN, we grouped the testing dataset into smaller bins sorted by event magnitude, source-receiver distance, and SNR.
The trained CNN is validated on these small testing datasets and its F-1 scores are plotted in Figure~\ref{fig:performance_analysis}.
The results generally follow our intuition: phases associated with events of larger magnitudes  (Figure~\ref{fig:f1_vs_mag}) and smaller distances (Figure~\ref{fig:f1_vs_dist}) being classified with higher accuracy.
Figure~\ref{fig:f1_vs_snr} demonstrates that the F-1 score is inversely proportional to the waveform SNR for both P and S phases.

\begin{figure}
    \centering
    \includegraphics[width=0.9\linewidth]{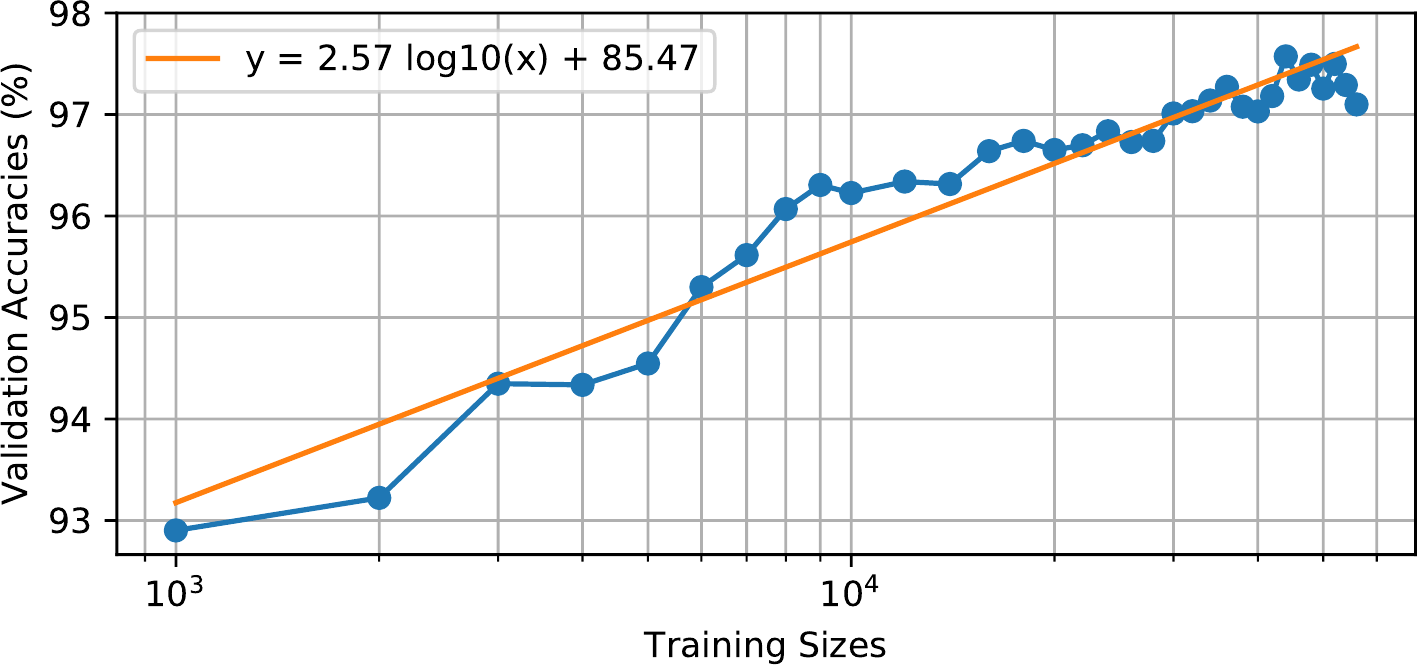}
    \caption{Validation accuracies vs. training dataset sizes (log scale) in blue. A line (log function) is fitted in orange.}
    \label{fig:accuracy-size}
\end{figure}

\paragraph{Flexible training set size}
As mentioned before, the overall 60,276 samples are split into training and validation datasets chronologically with different splitting ratios to explore the minimum required training dataset size.
Each split is trained up to 200 epochs and the model accuracy defined in \eqref{eq:accuracy} is shown in Figure~\ref{fig:accuracy-size}.
In general, the relationship between training set size and validation accuracy follows a log function as demonstrated %by the orange fitted log curve
in Figure~\ref{fig:accuracy-size}.
We note that \Acronym reaches 95\% accuracy with less than 6,000 training samples and 97\% with less than 30,000 training samples.
This largely reduces the amount of manual labeling needed to a reasonable level for practical applications.
For example, \Acronym only requires 300 manually picked aftershock events (for both P and S phases) per station on a 10-station network to achieve 95\% classification accuracy.

\paragraph{Fast deployment}
\Acronym is tested using the Nvidia GTX 1080 Ti GPU with 3,584 CUDA cores and 11\,GB memory.
The PyTorch machine learning package \citep{Paszke2017} and ObsPy seismic processing toolbox \citep{Beyreuther2010obspy} were used to automate our tests.
Online processing of one 20-sec window by the trained CNN  takes less than 0.3\,ms on average when feeding the input as 1000 windows per batch to exploit the maximum GPU memory size. % in our case.
This enables us to run the detector on the entire 31-day continuous 3-C waveforms recorded by 14 stations within two hours.
The time spent for phase picking depends on the number of detected phases and the merged window length.
In our study, it takes around 12 hours to pick all 30,000 catalog phases within the 31-day dataset.

\subsection{Event Detection on Continuous Waveforms}
With a 2-sec offset, the continuous waveforms are broken into a collection of 20-sec overlapped time windows for detection (see Figure~\ref{fig:flow-chart}).
\Acronym gives a label to each such 20-sec window as P phase, S phase, or noise.
Consecutive windows with the same label are merged into one longer window (Figure~\ref{fig:cpic_demo_detect}), e.g., four neighboring 20-sec windows expand to a 28-sec window.
As shown in Table~\ref{tab:performance_scores}, 98.6\% and 97.8\% of the catalog P and S phases are correctly detected (recall), while 97.0\% and 95.4\% detected P and S phases match a catalog phase (precision).

\begin{figure}[t!]
    \centering
    \includegraphics[width=\columnwidth]{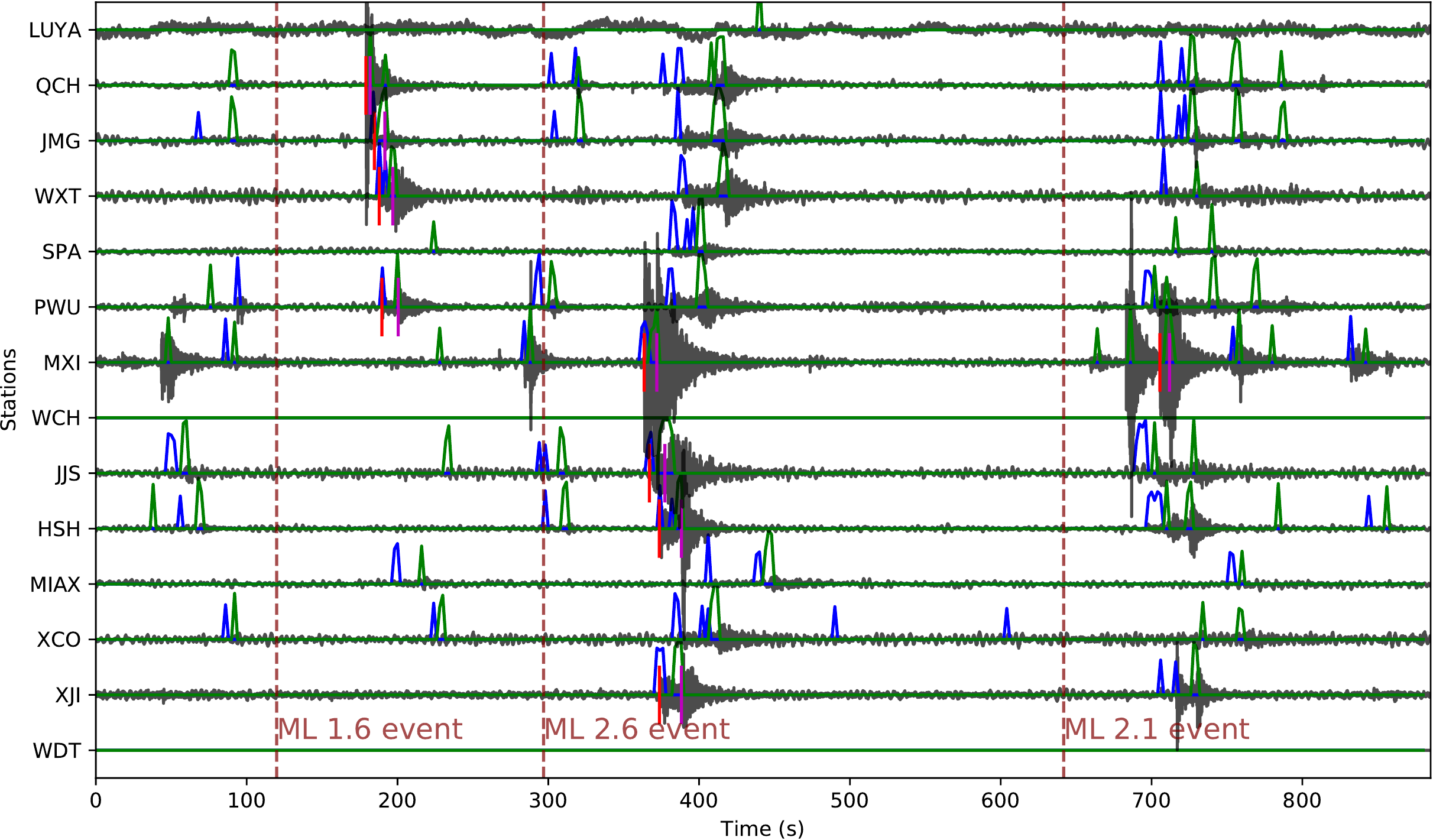}
    \caption{Detection example on 15-minute recording on 14 stations with three catalog events. Only vertical components are plotted. Blue and green curves show the probabilities of P and S phases. Red and magenta bars indicate the catalog P and S arrivals. Origin times of three catalog events are marked by the dashed vertical lines along with their magnitudes.}
    \label{fig:detection_example}
\end{figure}

Figure~\ref{fig:detection_example} shows the application of the \Acronym detector on a 15-minute continuous section across all 14 stations.
For the three catalog events (\ML{1.6}, 2.6, and 2.1, respectively), the \Acronym detector finds all phases picked in the catalog (marked by vertical bars in red for P phase and magenta for S phase).
Moreover, it detects additional phases for these three events on other stations that were missed by manual picking, e.g., P (blue peak) and S (green peak) phases around 400\,s on five additional stations (SPA, QCH, PWU, MIAX, and WXT) for the \ML{2.6} event.

On the other hand, additional phases are also detected, which might be associated with events missed in the catalog.
For example, two clusters of phases around 80\,s and 300\,s in Figure~\ref{fig:detection_example} exhibit reasonable moveout curves and may correspond to legitimate events.
To investigate these additional phase detections, we built a matched-filter (MF) enhanced catalog for one day (8/30/2008) following the procedure used by \cite{Meng2013}  (details explained in \ref{sec:mf}).
This MF catalog expands the original 150 events and 968 phases into 1,300 events and 12,200 phases for that day.
During the same time, \Acronym detects 4,123 seismic phases among which 2,892 (70\%) contain a phase in the MF catalog.
Further studies are needed to check whether the remaining 30\% correspond to actual events that are not similar to existing templates.

\begin{figure}[t!]
    \captionsetup[subfigure]{skip=-5mm,justification=justified,margin=10mm,singlelinecheck=false}
    \centering
    \begin{subfigure}{0.48\columnwidth}
        \centering
        \caption{}
        \label{fig:pick_error_dist_cpic_p}
        \includegraphics[width=\columnwidth]{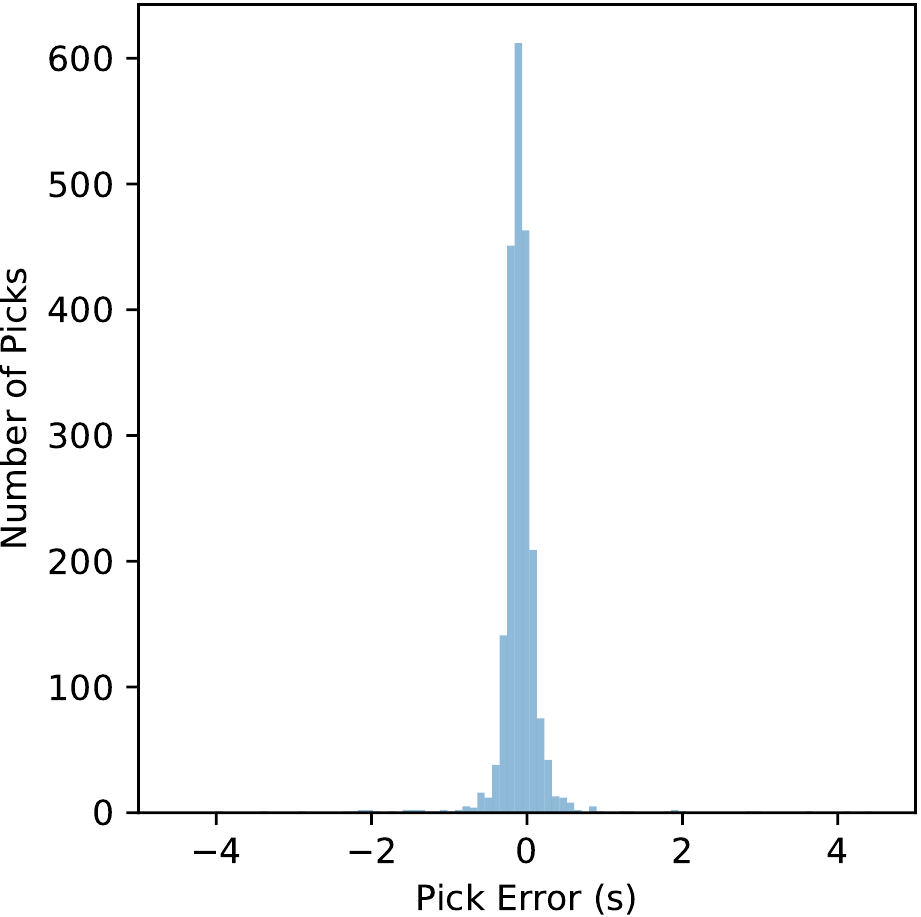}
    \end{subfigure}
    \hfill
    \begin{subfigure}{0.48\columnwidth}
        \centering
        \caption{}
        \label{fig:pick_error_dist_cpic_s}
        \includegraphics[width=\columnwidth]{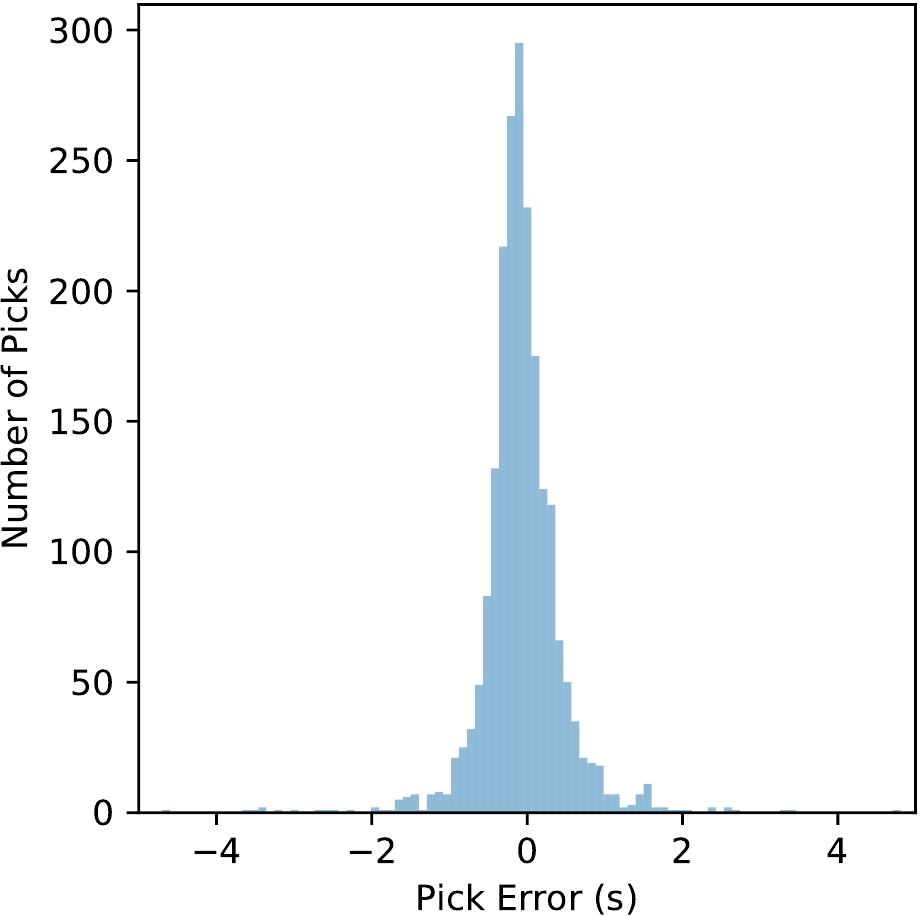}
    \end{subfigure}\\[3ex]
    
    \begin{subfigure}{0.48\columnwidth}
        \centering
        \caption{}
        \label{fig:pick_error_dist_ar_p}
        \includegraphics[width=\columnwidth]{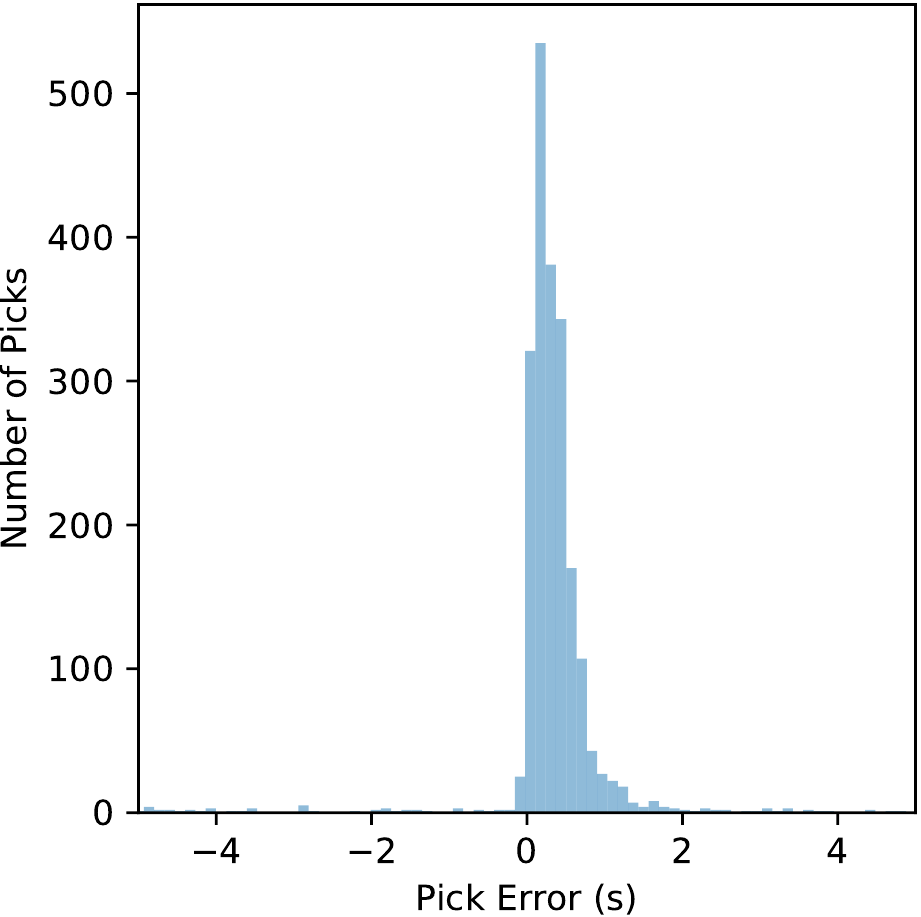}
    \end{subfigure}
    \hfill
    \begin{subfigure}{0.48\columnwidth}
        \centering
        \caption{}
        \label{fig:pick_error_dist_ar_s}
        \includegraphics[width=\columnwidth]{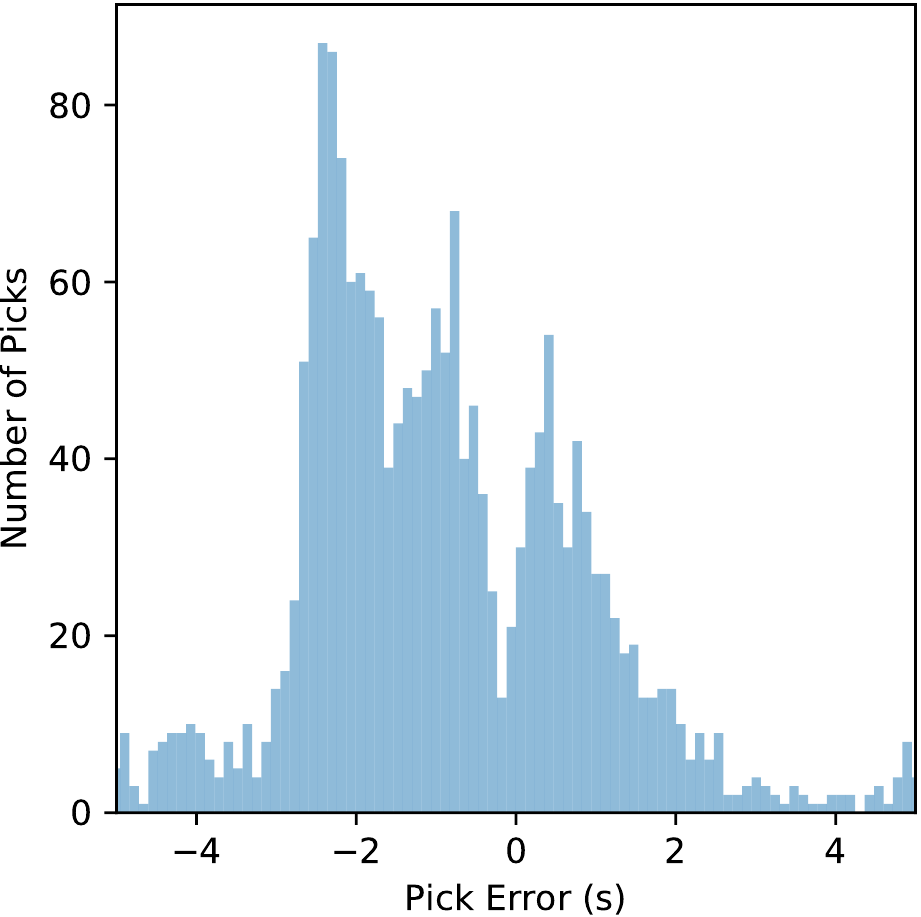}
    \end{subfigure}
    
    \caption{The distributions of picking errors ($E_\text{pick}$) of \Acronym (upper panels) and ObsPy AR picker (lower panels) on the validation dataset.}
    \label{fig:pick_error_dist}
\end{figure}
\begin{table}
    \centering
    \caption{Evaluation metrics for \Acronym and ObsPy AR picker on the validation dataset.}
    \label{tab:pick_results}
    \begin{tabular}{ l r r r r}
        \toprule
        Method &  $\mu(E_p)$ & $\mu(E_s)$ & $\sigma(E_p)$  & $\sigma(E_s)$ \\
        \midrule
        \Acronym picker (ms)& -79.0 & -78.9 & 138.8 & 293.0\\
        ObsPy AR picker (ms) & 311.4 & 936.3 & 671.6 & 1,697.0\\
        \bottomrule
    \end{tabular}
\end{table}
\subsection{Phase Picking on Catalog Events}
\label{sec:results-phasepicking}

\paragraph{Picking Results}
The detected windows are reprocessed by the CNN with a 0.1\,s offset to generate the \Acronym arrival times.
The picked arrival times are compared with the catalog phase arrivals and results from the ObsPy AR picker.
The error defined in \eqref{eq:error} is used to measure the performance of the P and S phase pickers separately.
Table~\ref{tab:pick_results} summarizes the statistics of picking errors for P and S phases from \Acronym and the ObsPy AR picker.
Errors for both P and S phases from \Acronym have much smaller standard deviations and biases than their counterparts from the ObsPy AR picker.
% with one-order-of-magnitude smaller bias.
Significant improvements are observed by applying \Acronym, especially for S-wave arrival times.
This is expected since picking S phase arrivals is more challenging for traditional methods due to interference from the P wave coda.
Figure~\ref{fig:pick_error_dist} compares the distributions of picking errors for P and S phases from \Acronym with the ObsPy AR picker.
The error distributions from both methods for P arrivals are narrower than those for S waves.
This is consistent with our intuition that P phase arrivals are clear and easier to pick.
Notice that both distributions from \Acronym are more symmetric than those from ObsPy AR picker.
\begin{figure}[t!]
    \captionsetup[subfigure]{skip=-1mm,justification=justified,margin=-2mm,singlelinecheck=false}
    \centering
    \begin{subfigure}{0.48\columnwidth}
        \centering
        \caption{}
        \label{fig:pick_example_good1}
        \includegraphics[width=\columnwidth]{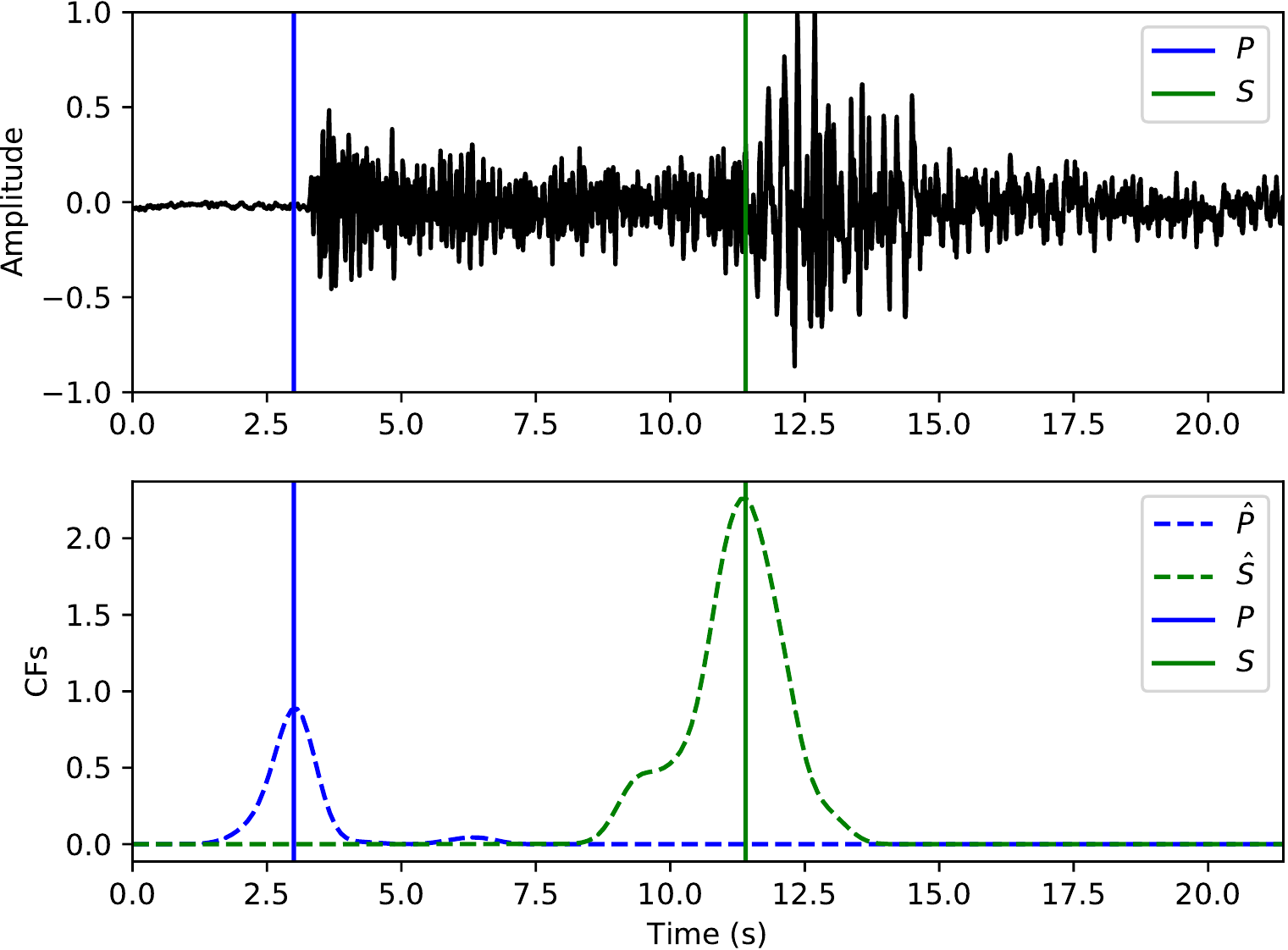}
    \end{subfigure}
    \hfill
    \begin{subfigure}{0.48\columnwidth}
        \centering
        \caption{}
        \label{fig:pick_example_good2}
        \includegraphics[width=\columnwidth]{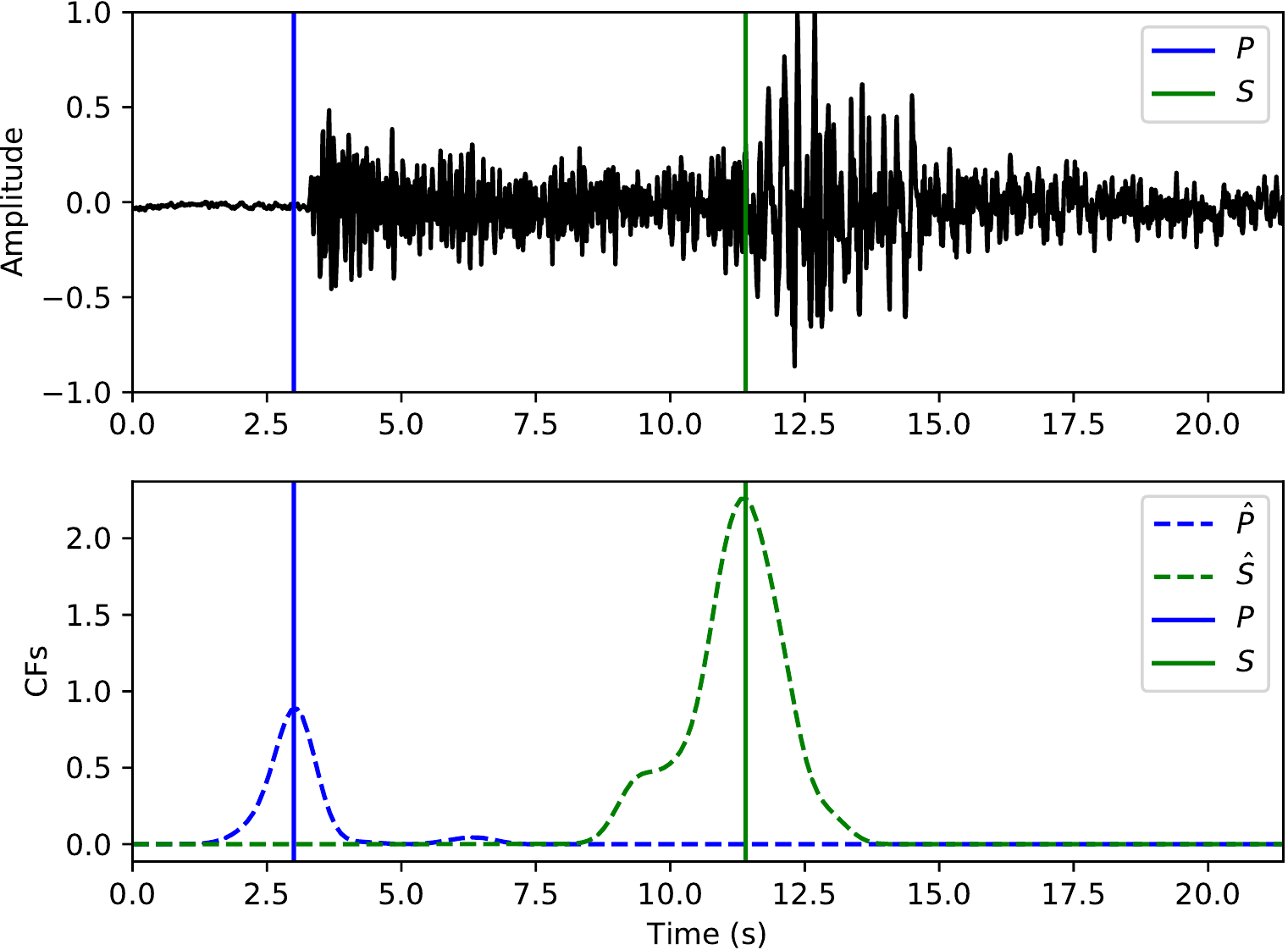}
    \end{subfigure}
    
    \begin{subfigure}{0.48\columnwidth}
        \centering
        \caption{}
        \label{fig:pick_example_good3}
        \includegraphics[width=\columnwidth]{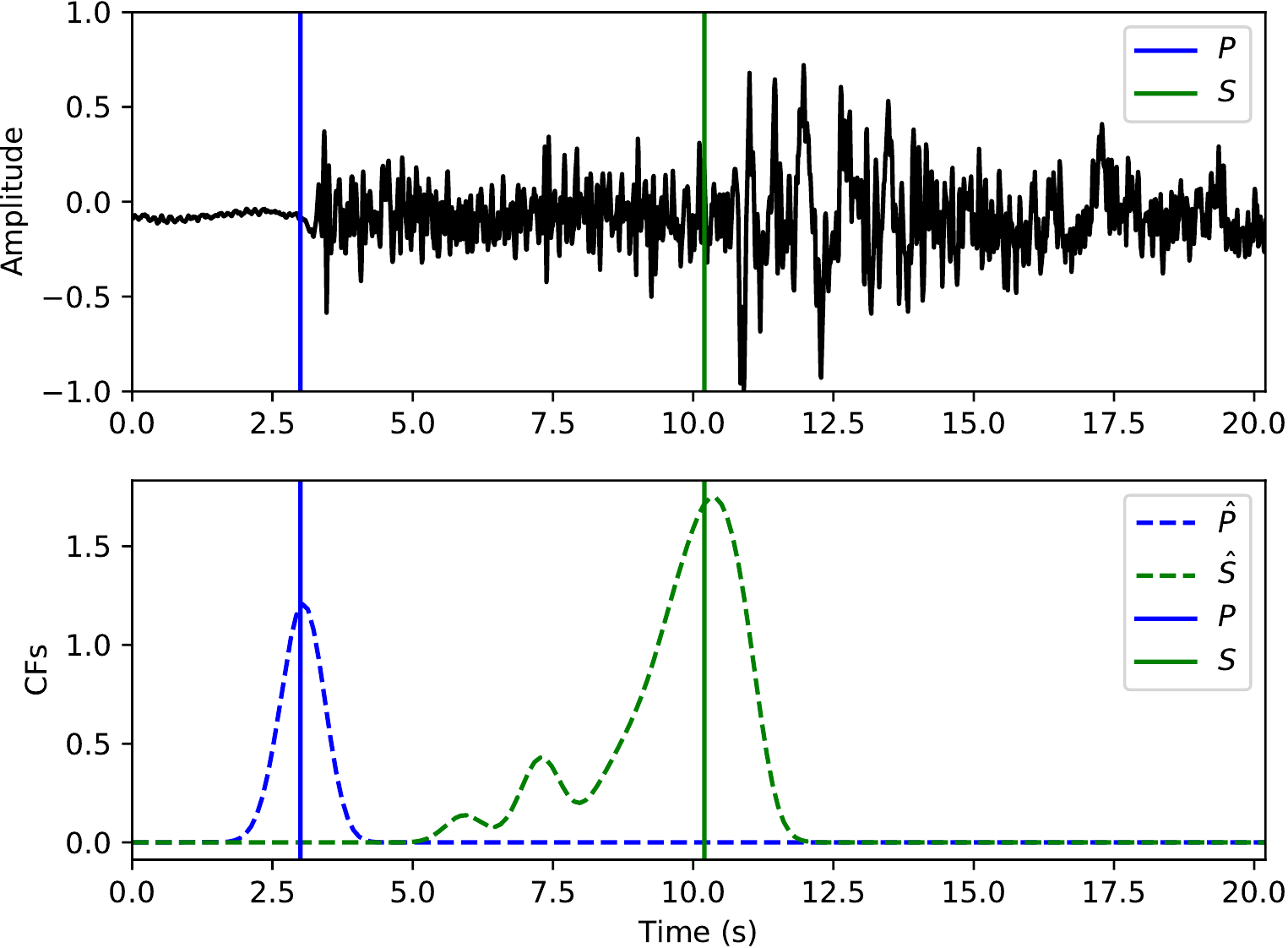}
    \end{subfigure}
    \hfill
    \begin{subfigure}{0.48\columnwidth}
        \centering
        \caption{}
        \label{fig:pick_example_good4}
        \includegraphics[width=\columnwidth]{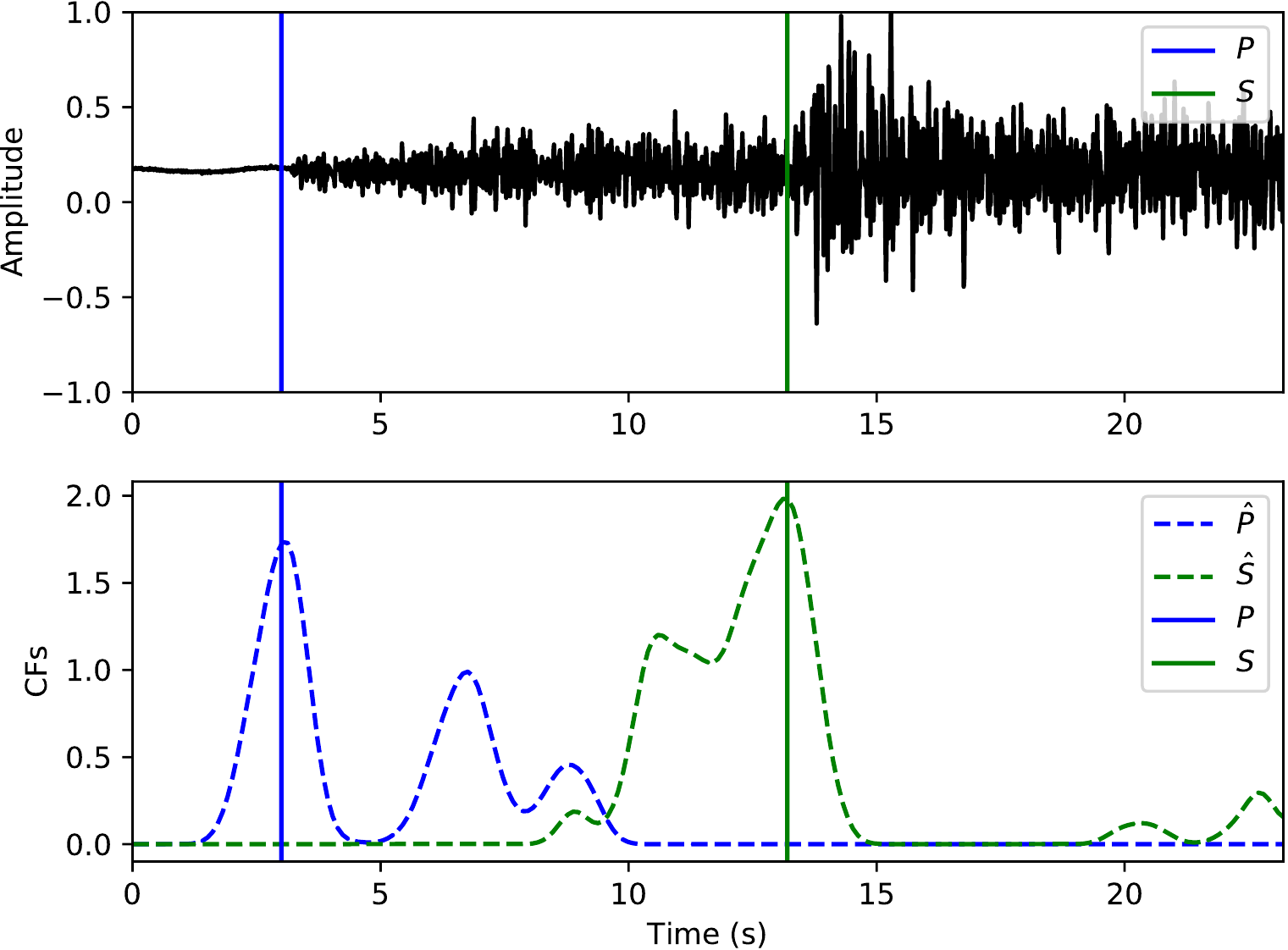}
    \end{subfigure}
    
    \begin{subfigure}{0.48\columnwidth}
        \centering
        \caption{}
        \label{fig:pick_example_good5}
        \includegraphics[width=\columnwidth]{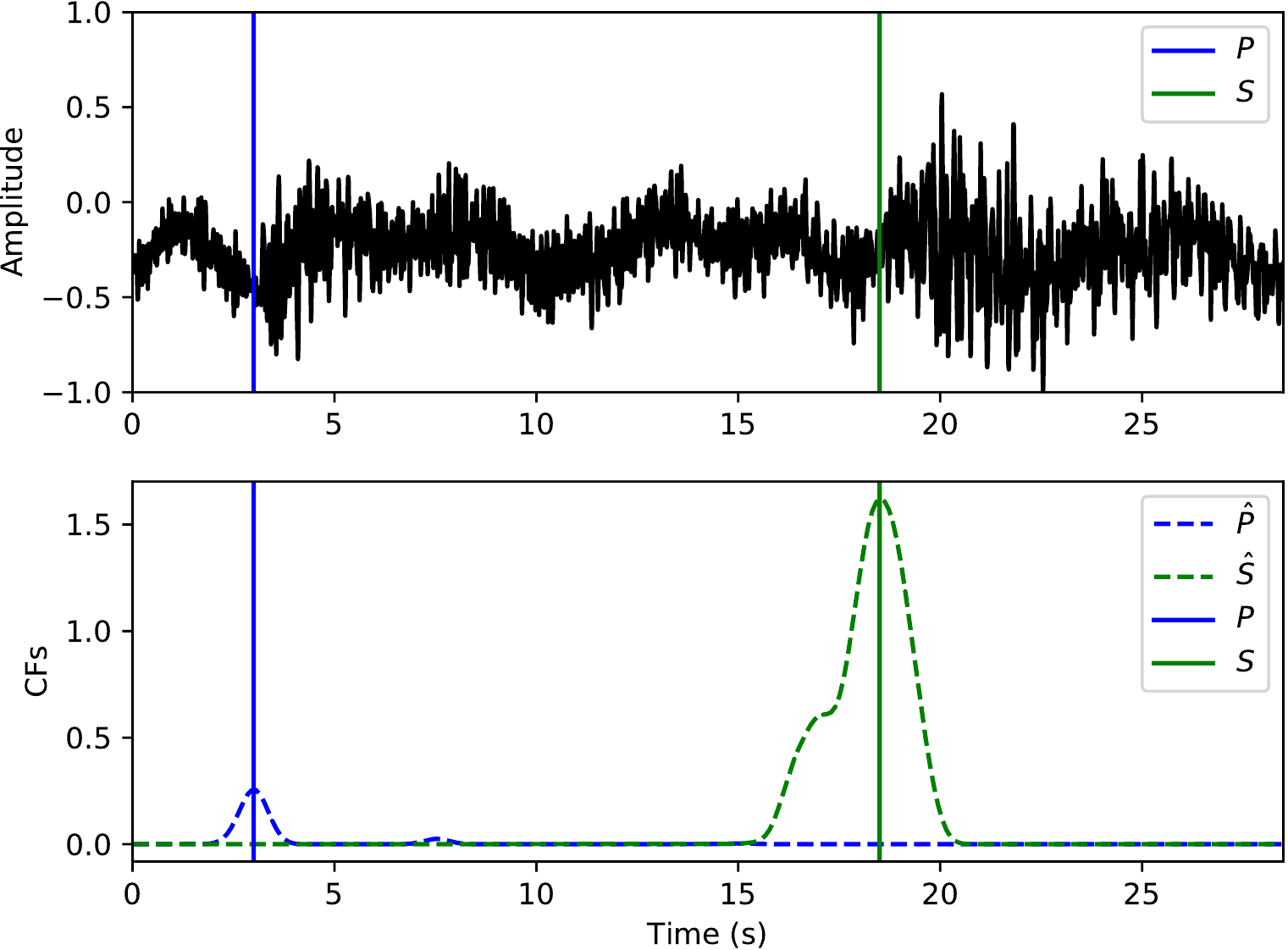}
    \end{subfigure}
    \hfill
    \begin{subfigure}{0.48\columnwidth}
        \centering
        \caption{}
        \label{fig:pick_example_good6}
        \includegraphics[width=\columnwidth]{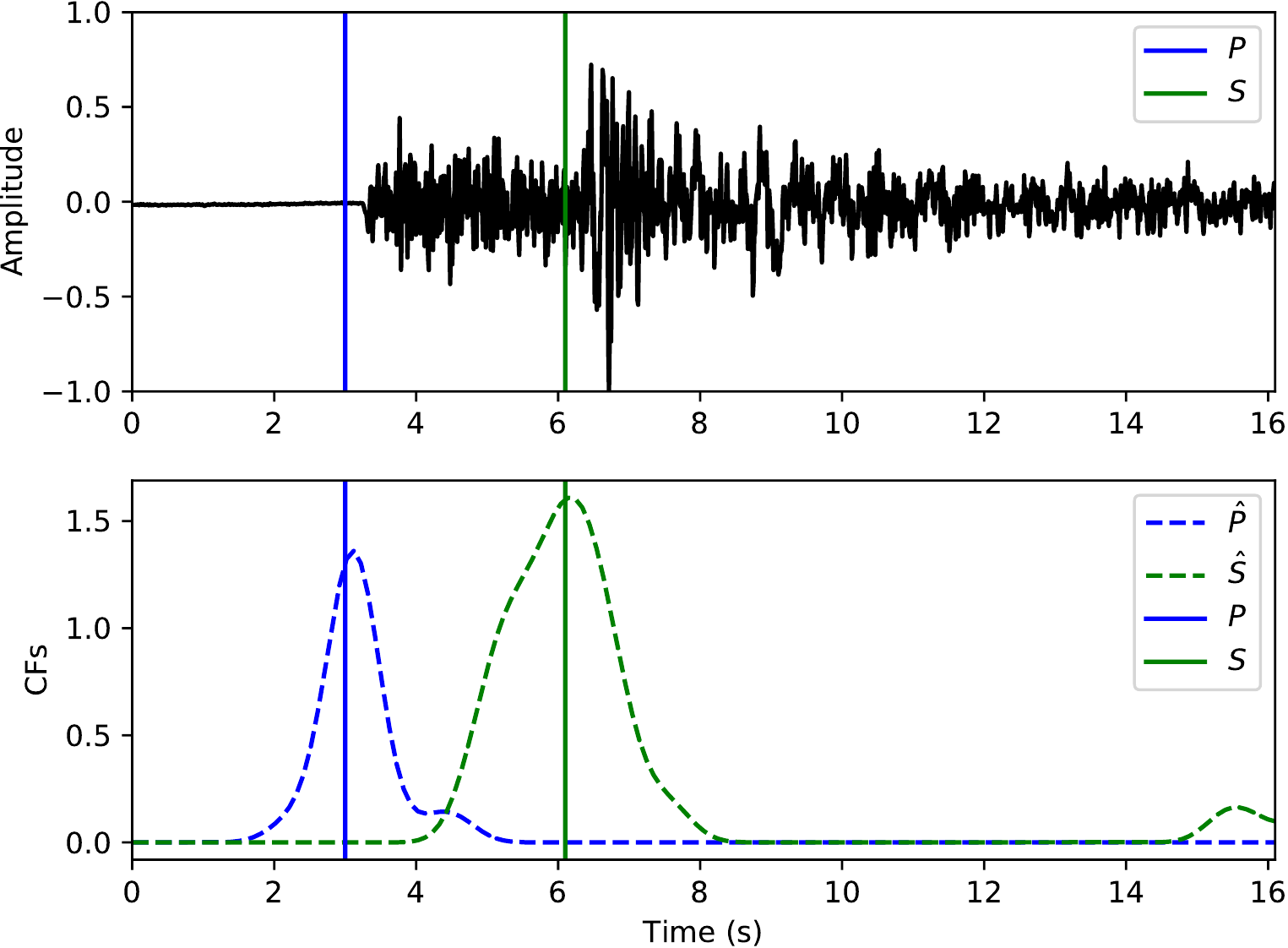}
    \end{subfigure}
    
    \caption{Examples of \Acronym picks that are consistent with manual picks. The upper panels of (a)\,--\,(f) are the vertical components from the 3-C waveforms used in the picking process.
        Vertical lines denote arrival-time picks.
        The lower panels show the characteristic functions (CFs) of $\hat{P}$ (blue) and $\hat{S}$ (green) used by \Acronym to pick the arrival times.
        %    Note that all three components are used to generate the corresponding CFs, but only the vertical components are shown.
    }
    \label{fig:pick_example_good}
\end{figure}
\begin{figure}[t!]
    \captionsetup[subfigure]{skip=-1mm,justification=justified,margin=-2mm,singlelinecheck=false}
    \centering
    \begin{subfigure}{0.48\columnwidth}
        \centering
        \caption{}
        \label{fig:pick_example_bad1}
        \includegraphics[width=\columnwidth]{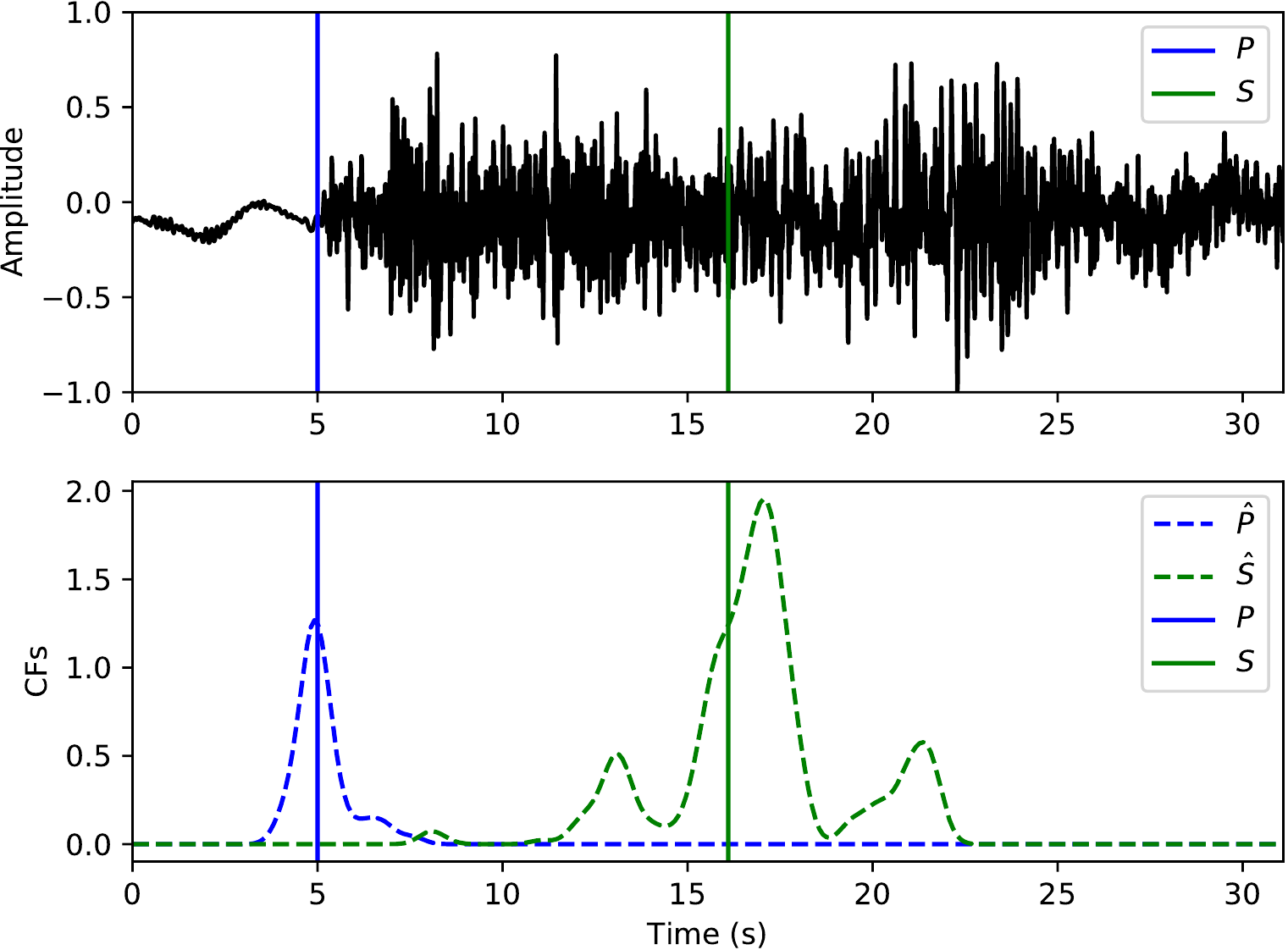}
    \end{subfigure}
    \hfill
    \begin{subfigure}{0.48\columnwidth}
        \centering
        \caption{}
        \label{fig:pick_example_bad2}
        \includegraphics[width=\columnwidth]{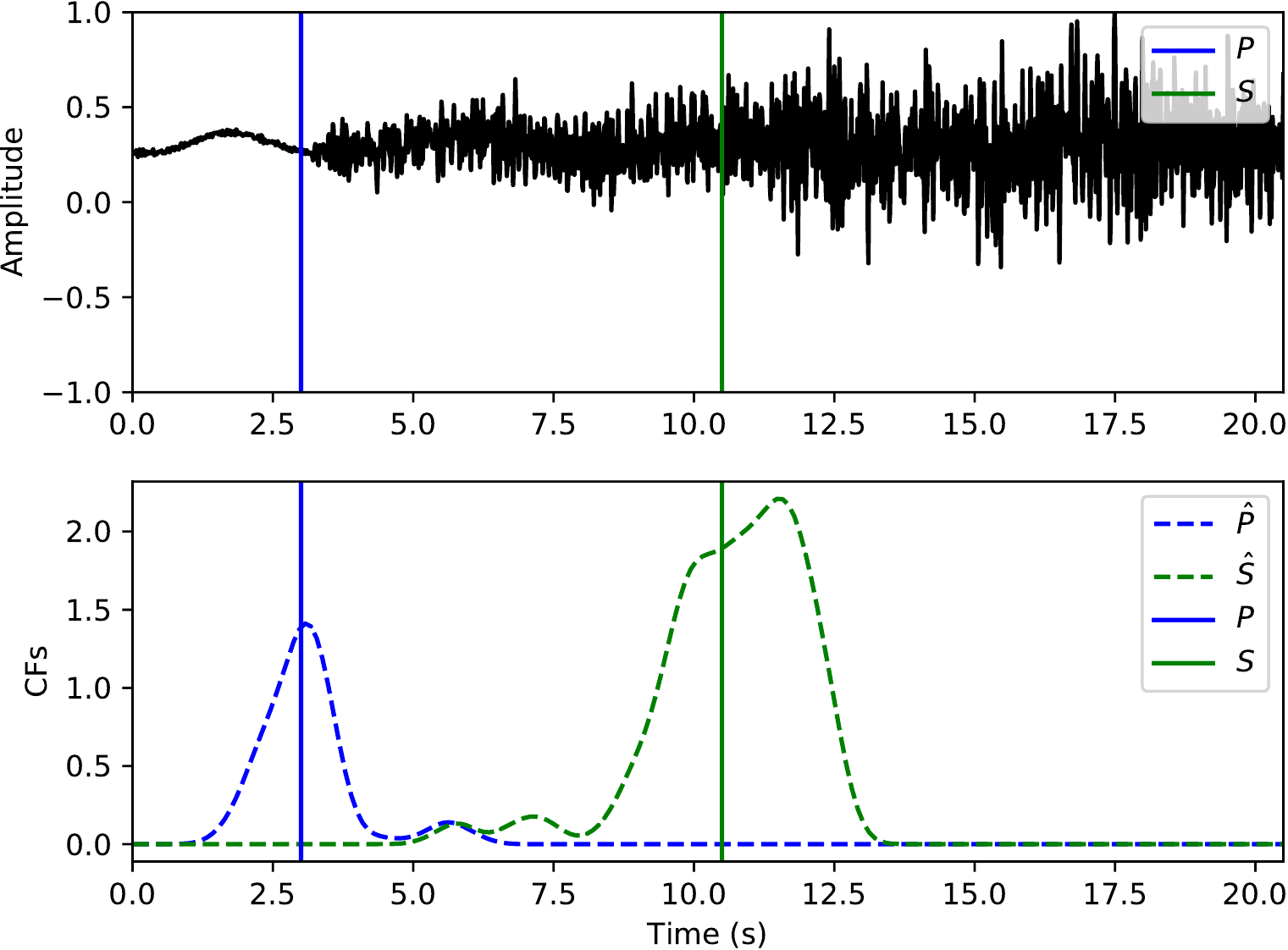}
    \end{subfigure}
    
    \begin{subfigure}{0.48\columnwidth}
        \centering
        \caption{}
        \label{fig:pick_example_bad3}
        \includegraphics[width=\columnwidth]{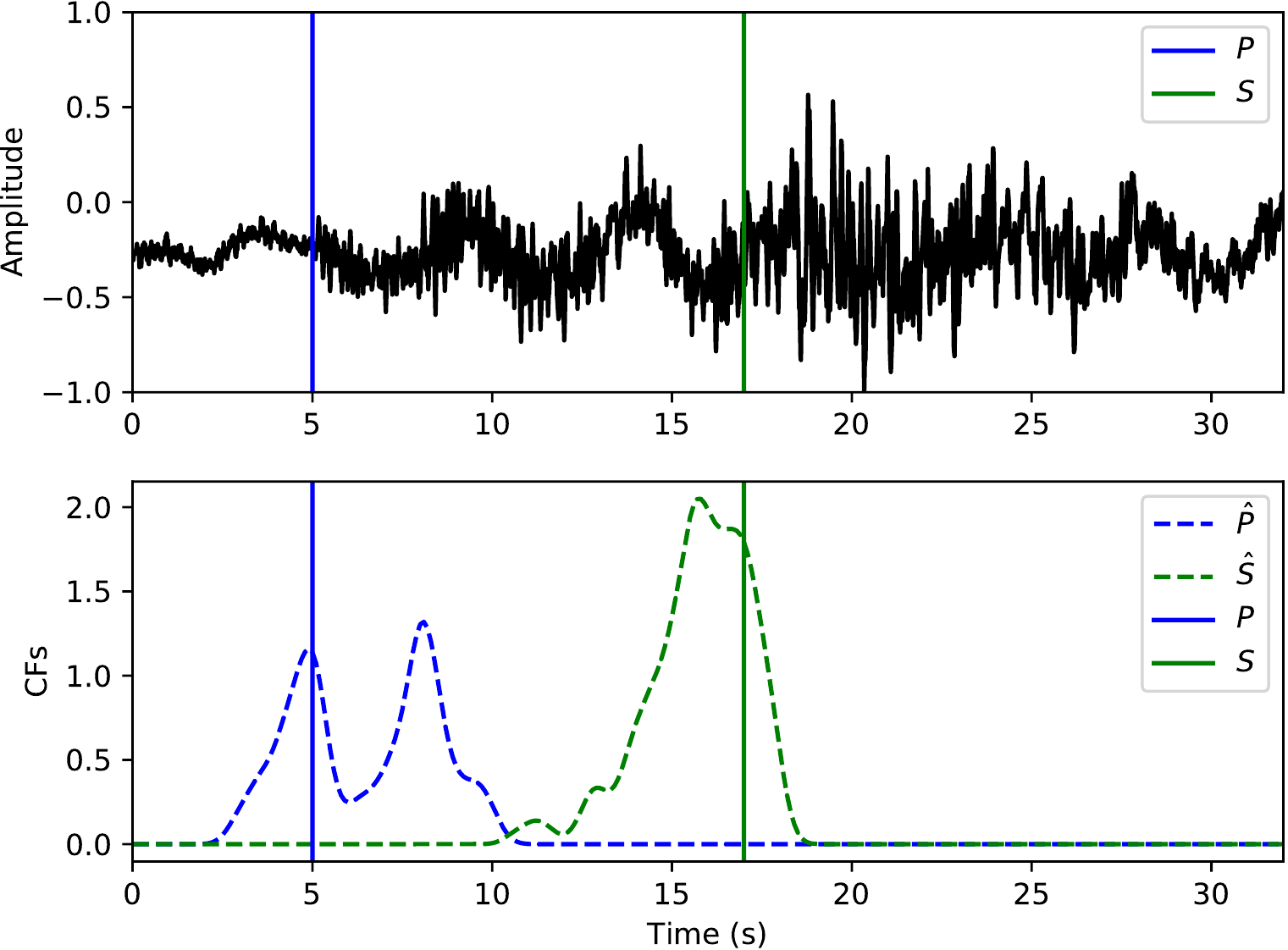}
    \end{subfigure}
    \hfill
    \begin{subfigure}{0.48\columnwidth}
        \centering
        \caption{}
        \label{fig:pick_example_bad4}
        \includegraphics[width=\columnwidth]{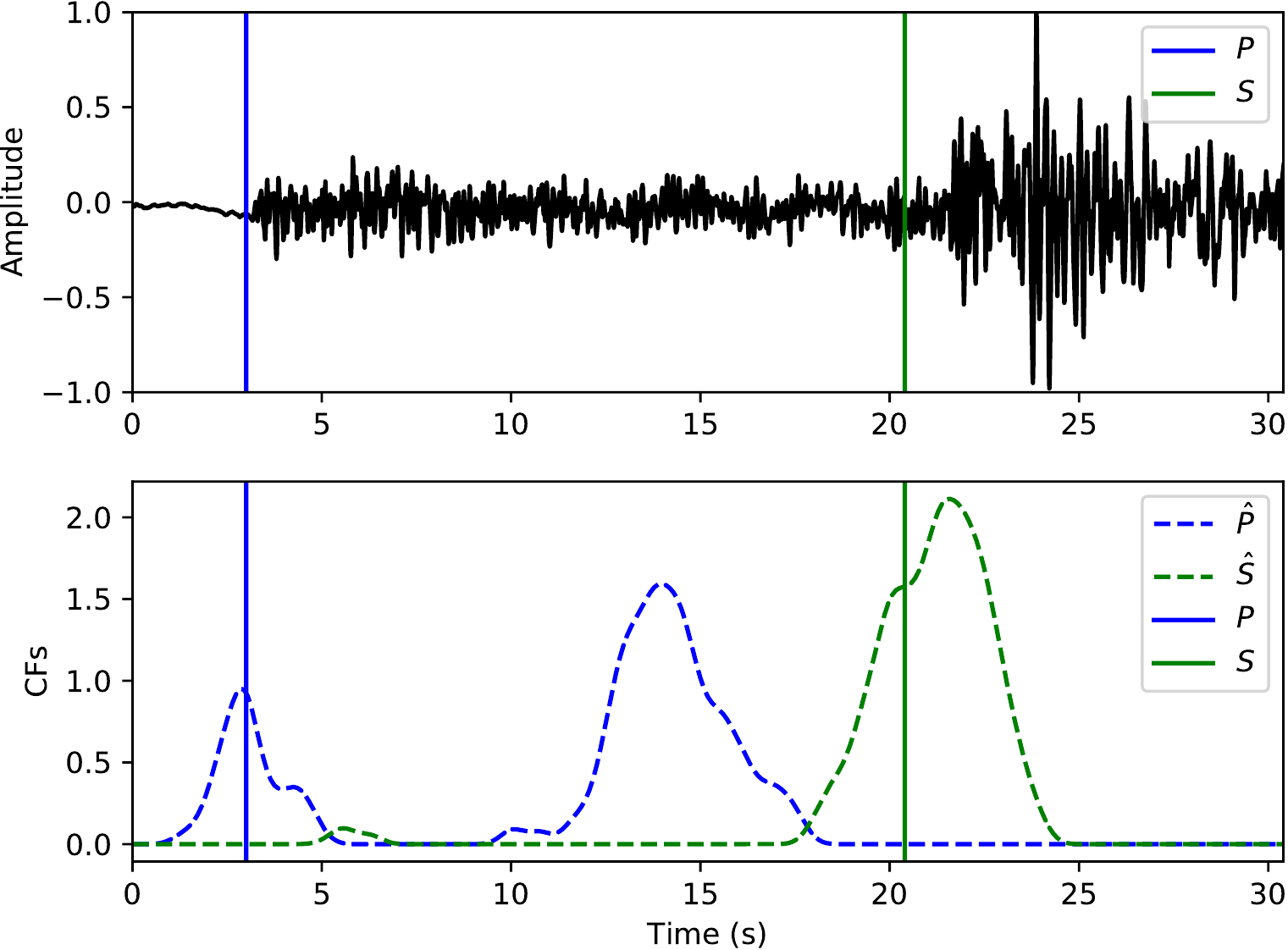}
    \end{subfigure}
    
    \begin{subfigure}{0.48\columnwidth}
        \centering
        \caption{}
        \label{fig:pick_example_bad5}
        \includegraphics[width=\columnwidth]{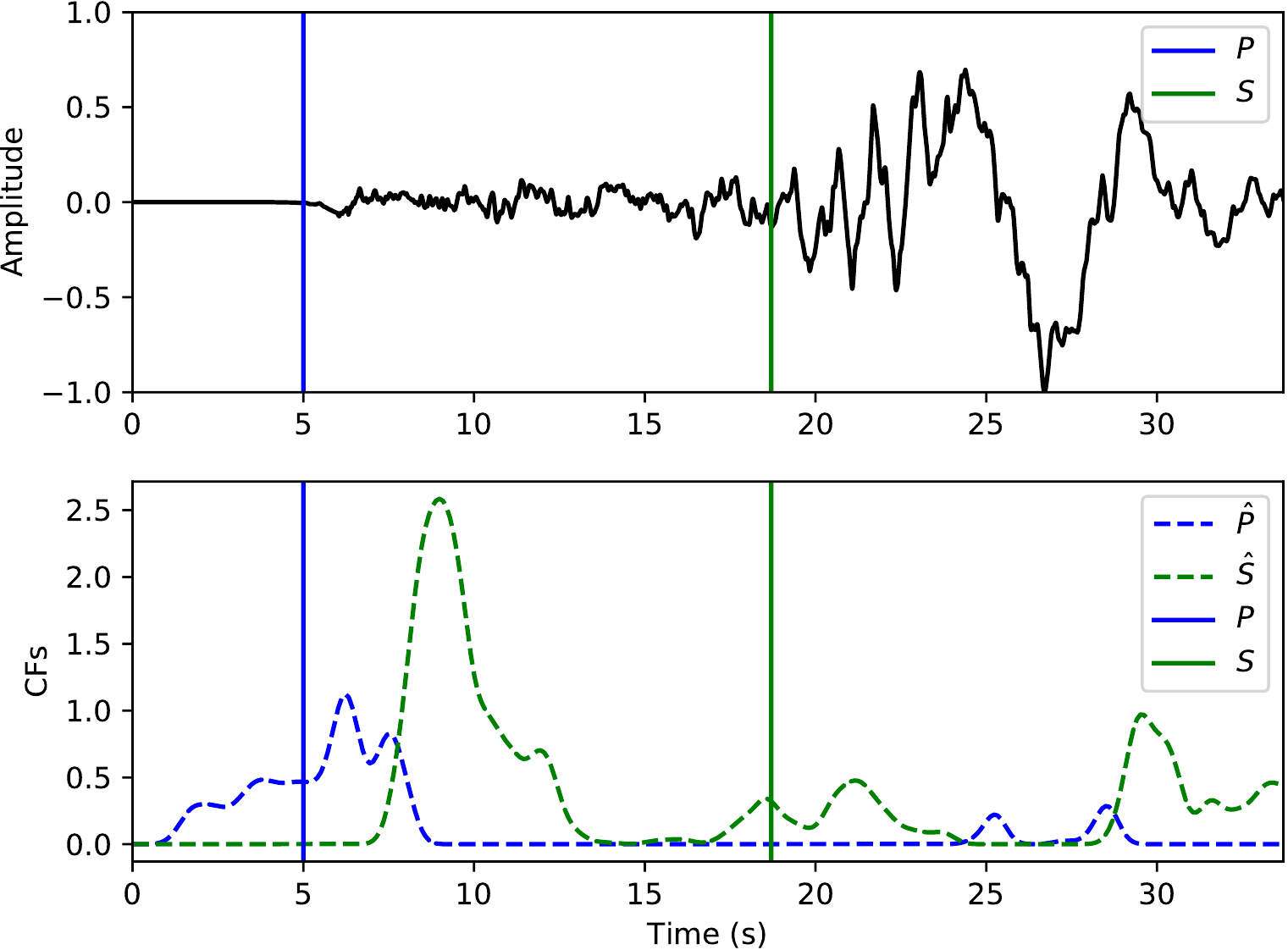}
    \end{subfigure}
    \hfill
    \begin{subfigure}{0.48\columnwidth}
        \centering
        \caption{}
        \label{fig:pick_example_bad6}
        \includegraphics[width=\columnwidth]{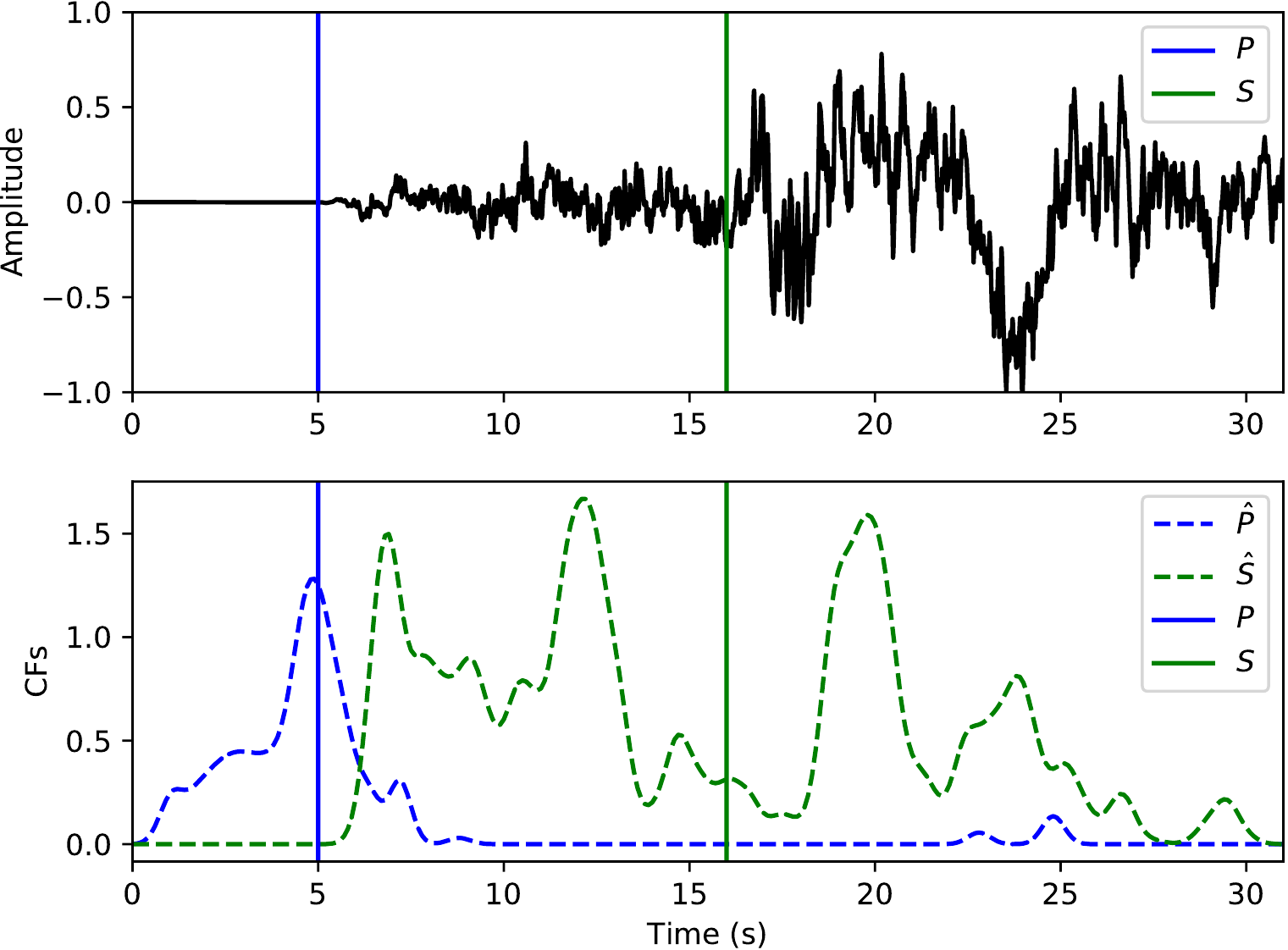}
    \end{subfigure}
    
    \caption{Examples of \Acronym picks that are inconsistent with manual picks. (a, b) are examples of ambiguous S picks. (c, d) are examples of multiple P picks. (e, f) are examples of an \ML{6.1} events on two distant stations. }
    \label{fig:pick_example_bad}
\end{figure}
\paragraph{Picking Examples}
Examples of arrival picking are given in Figure~\ref{fig:pick_example_good} and \ref{fig:pick_example_bad} to demonstrate \Acronym's performance.
Note that the waveforms displayed in the upper panels have mean removed and are scaled to have a maximum amplitude of one; however, the real inputs to the \Acronym model are the original raw waveforms.
Figures~\ref{fig:pick_example_good1} and \ref{fig:pick_example_good2} show the ideal cases where there is only one distinct peak in the CFs of both P and S phases that aligns perfectly with the catalog arrival times.
Multiple peaks are present in Figure~\ref{fig:pick_example_good3} and \ref{fig:pick_example_good4}, but the \Acronym picks correctly matched the manual picks.
Less ideal cases are shown in Figure~\ref{fig:pick_example_good5} and \ref{fig:pick_example_good6}  where \Acronym picks the correct arrival times but may have issues when the conditions are worse.
The noisy waveform in Figure~\ref{fig:pick_example_good5} results in a small peak for P wave around 3\,s, which may be buried under the noise floor if more severe noise were present.
\Acronym picked the arrival times in Figure~\ref{fig:pick_example_good6} correctly but has a small tail for the S phase at the end.
This small tail was successfully rejected due to its small amplitude, but it may become a false alarm if the relative peak amplitude of the S phase around 6\,s were much smaller.
This is also the case for Figure~\ref{fig:pick_example_good4}.
Examples of picks inconsistent with the catalog arrival times are also shown in Figure~\ref{fig:pick_example_bad}.
Unlike multiple peak cases shown in Figure~\ref{fig:pick_example_good}, the peak CFs from \Acronym in Figure~\ref{fig:pick_example_bad3} and \ref{fig:pick_example_bad4} is more than 1\,s from the manually picked arrivals.
Figure~\ref{fig:pick_example_bad5} and \ref{fig:pick_example_bad6} show incorrect picks of a \MW{6.1} event on two distant stations (SPA and WXT).
Since there are only two events with magnitude larger than \MW{6} in the given Wenchuan catalog, the trained model is ``inexperienced'' with such large events.
This is one of the disadvantages for training-based approaches: the model needs to see enough examples before it can provide reliable predictions.

\section{Discussion}
\label{sec:discuss}

In this study, we designed \Acronym to classify a 20-sec time window as noise, P phase or S phase based on training a CNN over a set containing 60,000 manually labeled windows.
The resulting classifier not only achieves more than 97\% accuracy for its original classification task but also serves as a key component for phase detection and picking.
The training process tweaks the weights of filters in the CNN model and reinforces the knowledge of seismic phase characteristics by iterative updates.
The resulting knowledge, encapsulated in the CNN representation of the continuous data, helps us to easily design a straightforward detection and picking system for seismic phases.
By using overlapping 20-sec windows with a fixed offset, the trained CNN provides a continuous output of probability values for its noise, P-phase, and S-phase classes.

\subsection{Comparison with other CNN approaches}
\label{sec:discuss-compare}

Another way to exploit deep learning for phase picking is to train the CNN for detection outputs and phase picking outputs directly.
As demonstrated in \citet{ZhuBeroza2018}, a likelihood function of seismic phases can be estimated for a given waveform instead of individual classification on each data point.
Trained on over a million labeled waveforms in Northern California (NCEDC 2014), PhaseNet \citep{ZhuBeroza2018} achieves better picking accuracy (51.5 vs. 138.8\,ms for P and 82.9 vs. 293.0\,ms for S).
However, we note that our dataset has not only more than one order-of-magnitude fewer labeled samples, but also challenging picking conditions\,--\,the benchmarks from the ObsPy AR picker have ten-times-larger standard deviation of picking errors.
As shown in Figure~\ref{fig:pick_error_dist_ar_p} and \ref{fig:pick_error_dist_ar_s}, the STA/LTA based AR-AIC picking method results in large uncertainty of the picked arrival times.
This is drastically different from the condition in \cite{ZhuBeroza2018} where the AR-AIC method results in picking errors with less than 200\,ms standard deviation.
Since our catalog is limited in the number of labeled waveforms and more challenging conditions, we elected to keep the picker simple and focus on the effectiveness of the CNN for feature extraction.

When comparing with \citet{Ross2018b}, the proposed CNN yields comparable detection accuracy (97.4\% vs. $>99$\%) even though it uses a relatively small training dataset (40,000 vs. $>1$ million training samples).
This is mainly because the task that the CNN classifiers are trained on is rather simple\,--\,the CNN easily extracts the key features that are needed to effectively separate the noise, P, and S phase windows from each other.
This agrees with our intuition and the role of human analysts: noise, P phase, and S phase are very distinctive in good SNR cases.
Just as analysts learn to pick correct seismic phases by looking at examples of P and S phases, our CNN classifiers are trained on good SNR cases labeled by manual picking.
Compared to traditional methods, the CNN can be applied quickly and automatically to a large volume of data with more challenging conditions, such as variable SNR.

\subsection{\Acronym applied to induced earthquake dataset in Oklahoma, USA}
\label{sec:discuss-ok}

\begin{figure}
    \centering
    \includegraphics[width=0.7\columnwidth]{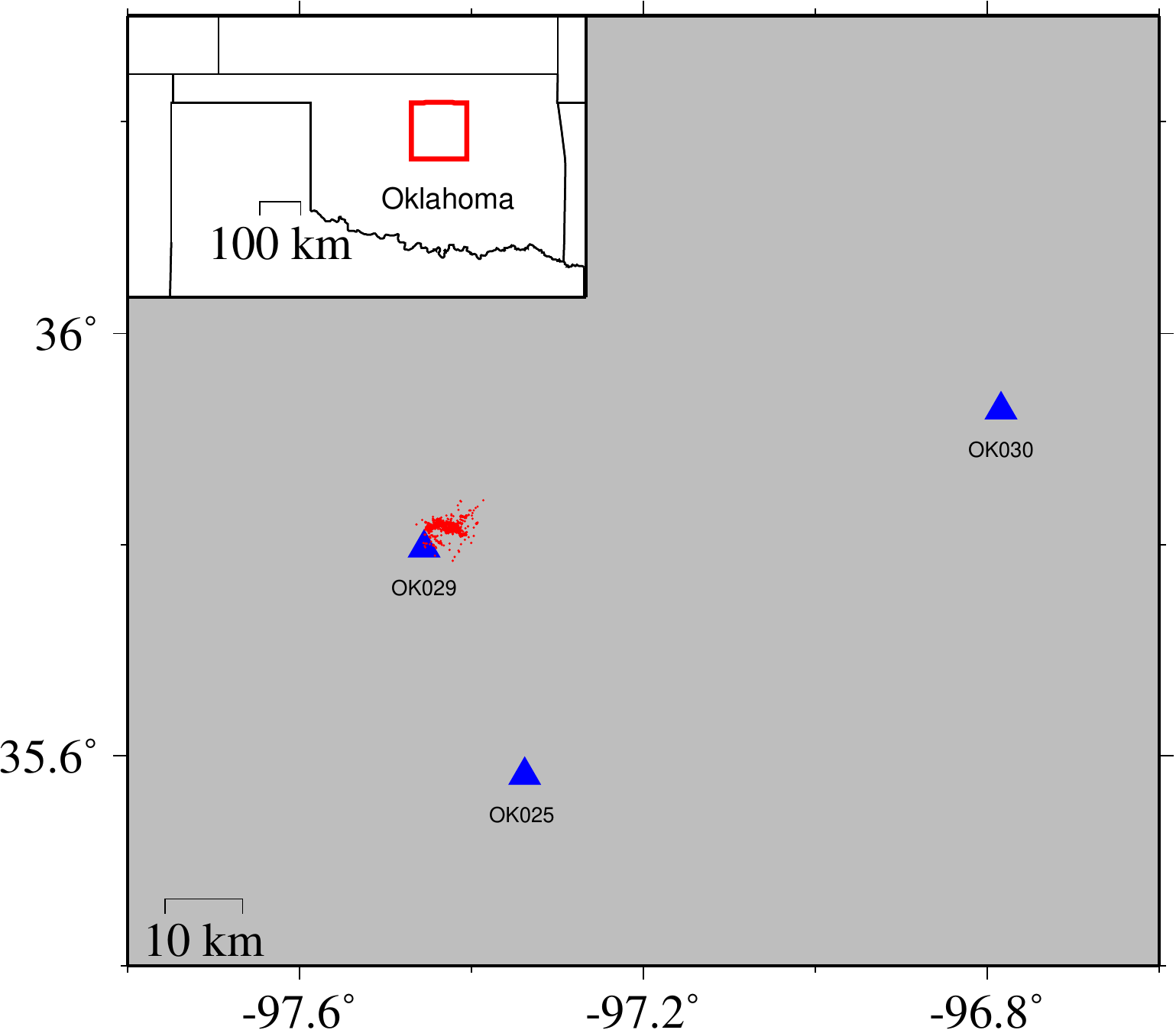}
    \caption{Map of study region in Oklahoma, central U.S. Red dots are 890 events with P and S phase arrivals and blue triangles are broadband stations of the US Geological Survey Network (GS). }
    \label{fig:ok-map}
\end{figure}

To validate how well \Acronym generalizes to another dataset, we apply the CNN trained on aftershocks in Wenchuan, China to a dataset containing likely human-induced earthquakes in Oklahoma (OK), USA \citep{Chen2018}.
As shown in Figure~\ref{fig:ok-map}, 890 events were manually picked with P and S phases on three stations (OK025, OK029, and OK030).
This results in a small catalog dataset with approximately 5,000 labeled samples.
When we applied the original CPIC classifier trained on the Wenchuan dataset, it achieved accuracy above 90\% on the two near stations (OK025 and OK029), but not on the far station (OK030) as shown in Table~\ref{tab:ok-results}.

\begin{table}
    \centering
    \caption{CPIC accuracy when testing on a three-station seismic dataset in OK, USA. The first row shows the performance of directly applying CPIC as trained on the Wenchuan, China dataset,  while the second row shows the enhanced accuracy after fine-tuning CPIC on 2,000 training samples from the Oklahoma region.}
    \label{tab:ok-results}
    \begin{tabular}{l c c c c}
        \toprule
        Station & OK025 & OK029 & OK030 & All\\
        \midrule
        Original (\%) & 95.7 & 92.2 & 69.9 & 87.5\\
        Fine-tuned (\%) & 98.8 & 96.2 & 94.2 & 97.0\\
        \bottomrule
    \end{tabular}
\end{table}

Next, we retrained the model by fine-tuning only the fully-connected (FC) layer that classifies feature vectors into probabilities of phase/noise classes; the 11 convolutional layers were kept fixed.
After fine-tuning the classifier on approximately 2,000 samples ($\approx 350$ events), the accuracy on all three stations is above 94\% with an overall accuracy at 97.0\%.
This shows that the convolutional layers in the \Acronym model capture the essential representation of a seismic wave needed for phase classification.
After fine-tuning the classification layer (FC), the \Acronym model trained on one region can be generalized to other regions for different event types (aftershocks vs.\ induced earthquakes).

\section{Conclusions}
In this and other recent studies, CNNs have shown clear potential for efficiently processing large volumes of seismic waveform data with accurate results.
Usually, CNN-based approaches require a large training dataset with accurate labels, provided by human analysts.
In this paper, we demonstrated an alternative path when using deep learning for seismic processing.
Instead of designing and training a CNN to accomplish the phase detection and picking tasks directly, we trained a CNN-based classifier that categorizes a seismic window into three classes: P, S, or Noise.
This allows us to train a relatively simple CNN with a smaller training set.
The detection and picking task is then accomplished by repeatedly applying the classifier on overlapping windows from continuous waveforms.

We named this processing framework \Acronym and tested it on 3-C data collected from the aftershock zone of the 2008 \MW{7.9} Wenchuan earthquake.
\Acronym achieves over 97.5\% phase detection rate while finding a significant number of potential phases missed by manual picking.
\Acronym also has a phase picking accuracy for which almost all of its picks are within $\pm300$\,ms of the manually labeled picks (Figure~\ref{fig:pick_error_dist}).
More importantly, \Acronym's processing time is remarkably small: on a desktop workstation with an Nvidia GTX1080 Ti GPU, it takes 2\,hrs to detect and 12\,hrs to pick phases on 3-C continuous data recorded for 31 days on 14 stations.
When compared to an expanded catalog for one day, the aggregation of picks by \Acronym on all stations detects all events found by manual picking and finds additional events missed by manual picking.
Furthermore, 70\,\% of the picks from \Acronym can be confirmed by a matched filter enhanced catalog.
The trained model also reached 97\% accuracy on a dataset from a different region after fine-tuning one layer of the model on a small training set.
%With its small training set requirement and fast application on single-station waveforms, \Acronym is a promising deep learning solution for various phase detection and picking tasks.
Thus \Acronym has the potential to be applied to regions where manual pickings are sparse, but a large volume of unpicked waveforms is available.

\newpage
\bibliography{pepi.bib}

\newpage
\appendix

\setcounter{figure}{0} 
\setcounter{table}{0}

\newpage
\section{Window length}
\label{sec:window_length}

\begin{figure}[b!]
    \centering
    \includegraphics[width=\linewidth]{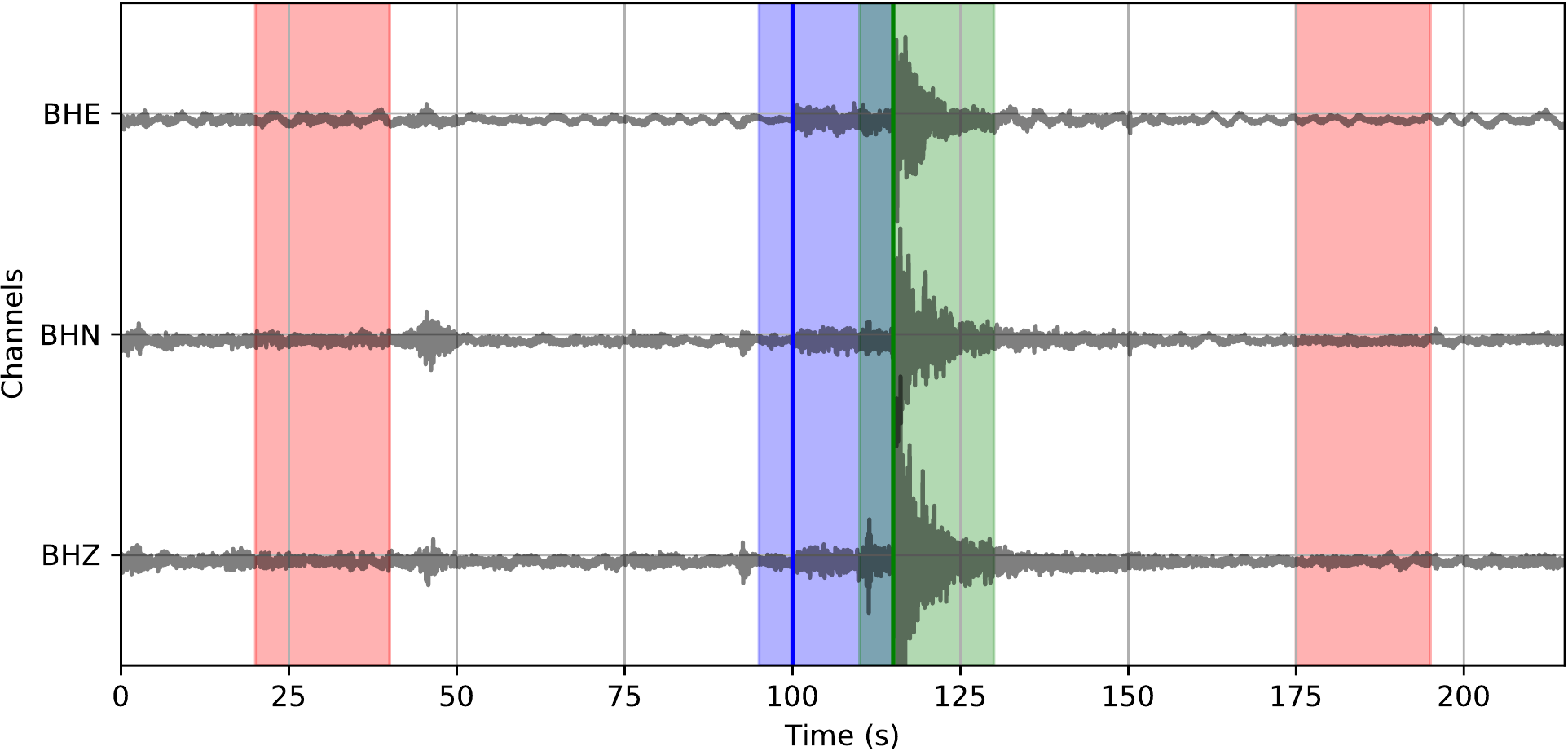}
    \caption{An example of three-component seismogram recorded at station HSH from which 20-sec long time windows are extracted for both P (blue) and S (green) phases. Noise (red) windows are cut one-minute before P and after S phases. Sampling rate is 100\,Hz. The arrival times of P and S phases are marked by vertical blue and green solid lines, respectively.}
    \label{fig:wenchun-windows}
\end{figure}

For each manually picked phase, we define a 20-sec long window starting 5\,s before the pick and ending 15\,s after as one window of a seismic phase (Figure~\ref{fig:wenchun-windows}).
A long time window was chosen so that there is a high likelihood that a P-wave window contains some S-wave at its end and that S-wave windows contain some P-wave coda at the beginning.
This window definition implicitly embeds the normal sequential relationship between P and S wave phases in the labeled dataset itself.
As shown in Table~\ref{tab:window-length}, some other typical windows lengths were tested, and those larger than 10\,s worked better for this dataset.

\vspace{1cm}
\begin{table}[h]
    \centering
    \caption{Classifier accuracy (defined in \eqref{eq:accuracy}) vs.  window lengths.}
    \label{tab:window-length}
    \begin{tabular}{l c c c c c}
        \toprule
        Window Length (sec) & 2.5 & 5 & 10 & 20 & 40\\
        \midrule
        Accuracy(\%) & 94.7 & 96.3 & 96.9 & 97.4 & 97.2\\
        \bottomrule
    \end{tabular}
\end{table}

\newpage
\section{Pre-processing}
\label{sec:preprocessing}

\begin{figure*}[h]
    \captionsetup[subfigure]{skip=-0.5mm}
    \centering
    \begin{subfigure}{\linewidth}
        \centering
        \includegraphics[width=\columnwidth]{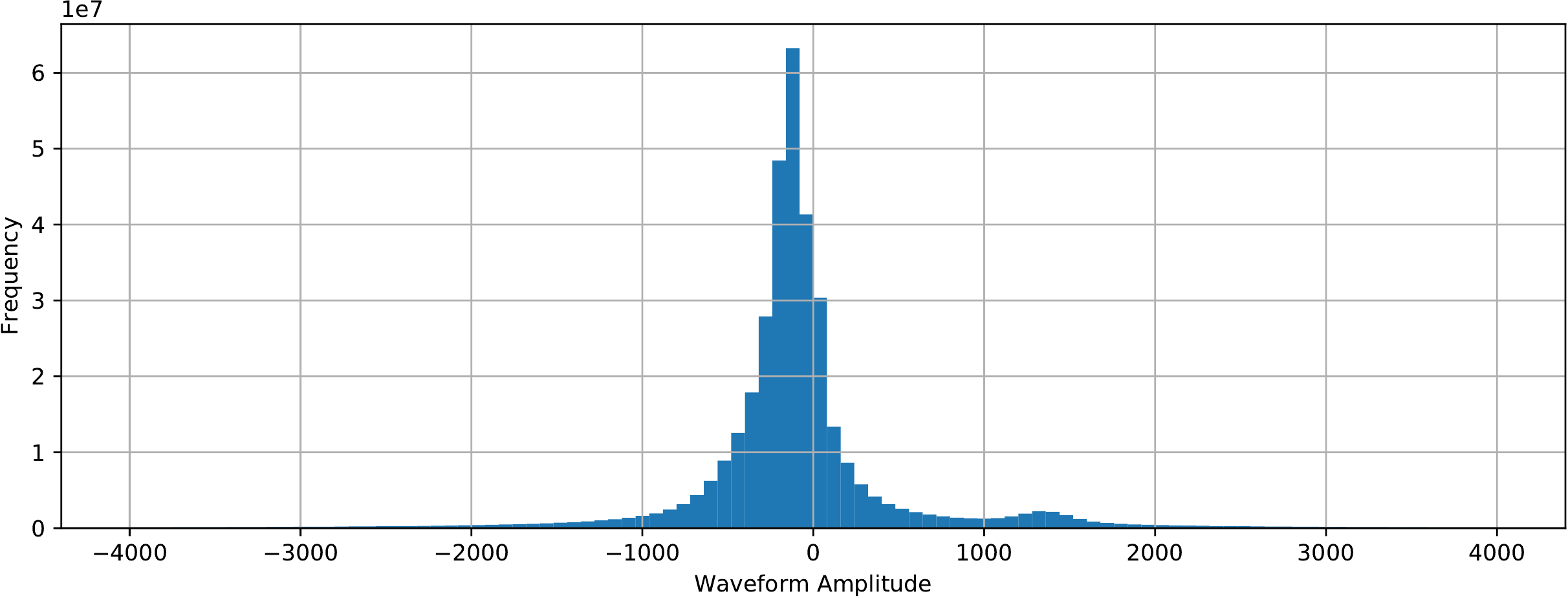}
        \caption{Distribution of waveform amplitude.}
        \label{fig:amplitude_dist}
    \end{subfigure}\\[2ex]
    \begin{subfigure}{0.43\linewidth}
        \centering
        \includegraphics[width=\linewidth]{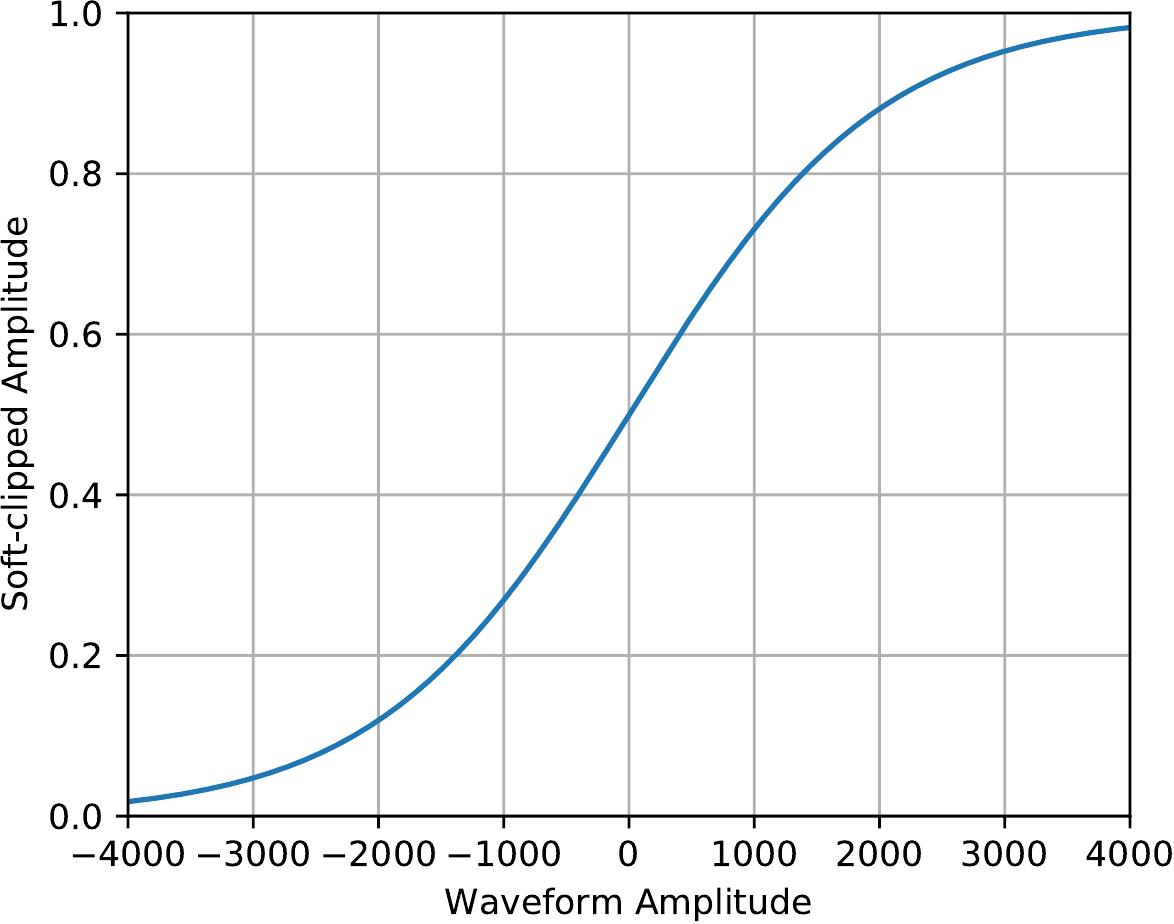}
        \caption{Logistic Function with $k=0.001$.}
        \label{fig:logistic_function}
    \end{subfigure}
    \hfill
    \begin{subfigure}{0.52\linewidth}
        \centering
        \includegraphics[width=\linewidth]{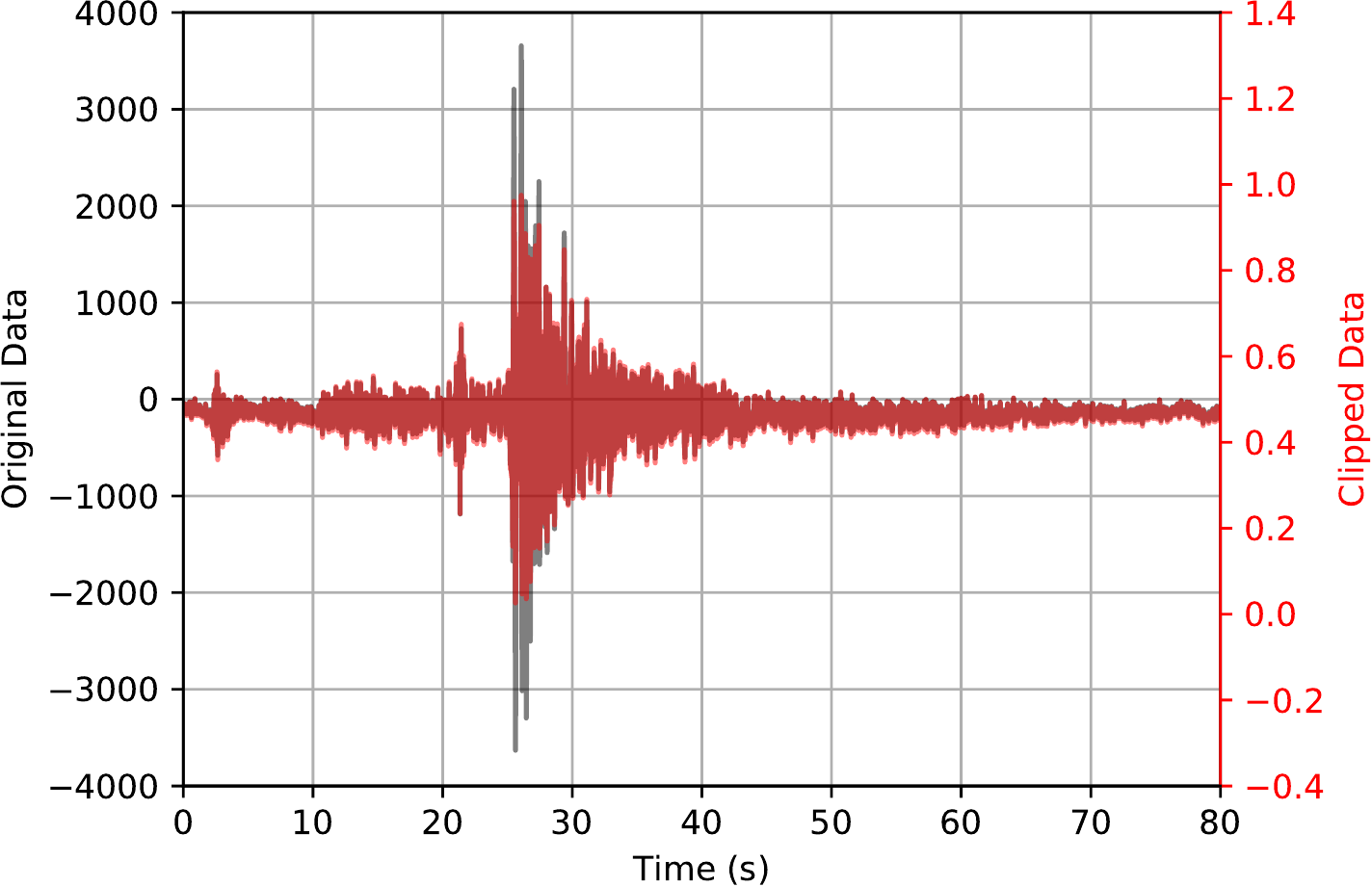}
        \caption{Soft clipping effect}
        \label{fig:soft_clip}
    \end{subfigure}
    \caption{Preprocessing for \Acronym: (a) waveform amplitude distribution; (b) soft clipping with a logistic function on the input data and (c) example of soft-clipped signal. Note that large amplitude signals in the original input (black) are reduced significantly on the clipped signal (red) while the small amplitude part is unchanged.}
\end{figure*}

Minimal preprocessing steps are performed on the raw seismic waveform in order to explore the limitations of ``expressiveness'' of the CNN.
It is believed that a sufficiently complex CNN can take the necessary data manipulation, such as band-pass filtering, into account if it is learned to be significant to the final classification task.
\paragraph{Soft-clipping method}
On the other hand, we observed that the dynamic ranges of the labeled events vary dramatically from each other.
This may result in the masking of weak events by stronger ones due to their amplitude difference.
Moreover, higher precision may be required after batch normalization due to such differences.
Since the GPU we used in this study works more efficiently for single-precision floating-point numbers, the dynamic range also imposes a hardware challenge.
Hence, we apply a soft clipping process based on a logistic function, which is shown in Figure~\ref{fig:logistic_function},
\begin{equation}
    f(x) = {1}/{(1 + e^{-kx})}
    \label{eq:expit}
\end{equation}
where $x$ is the original amplitude, and $k$ is chosen empirically based on the maximum amplitude in the original signal.

The soft clipping process, which is applied to all labeled data and continuous data with the same $k$ value, keeps the input data range between 0 and 1, as well as reducing the relative amplitudes of strong and weak events.
Figure~\ref{fig:soft_clip} illustrates that the soft-clipping process only suppresses the large amplitude signal while keeping the small one unchanged.
Figure~\ref{fig:amplitude_dist} shows that the amplitude of most traces is less than $4000$, thus we chose $k=0.001$ and the resulting soft-clipping function is shown in Figure~\ref{fig:logistic_function}.

\begin{figure}[h!]
    \centering
    \includegraphics[width=\linewidth]{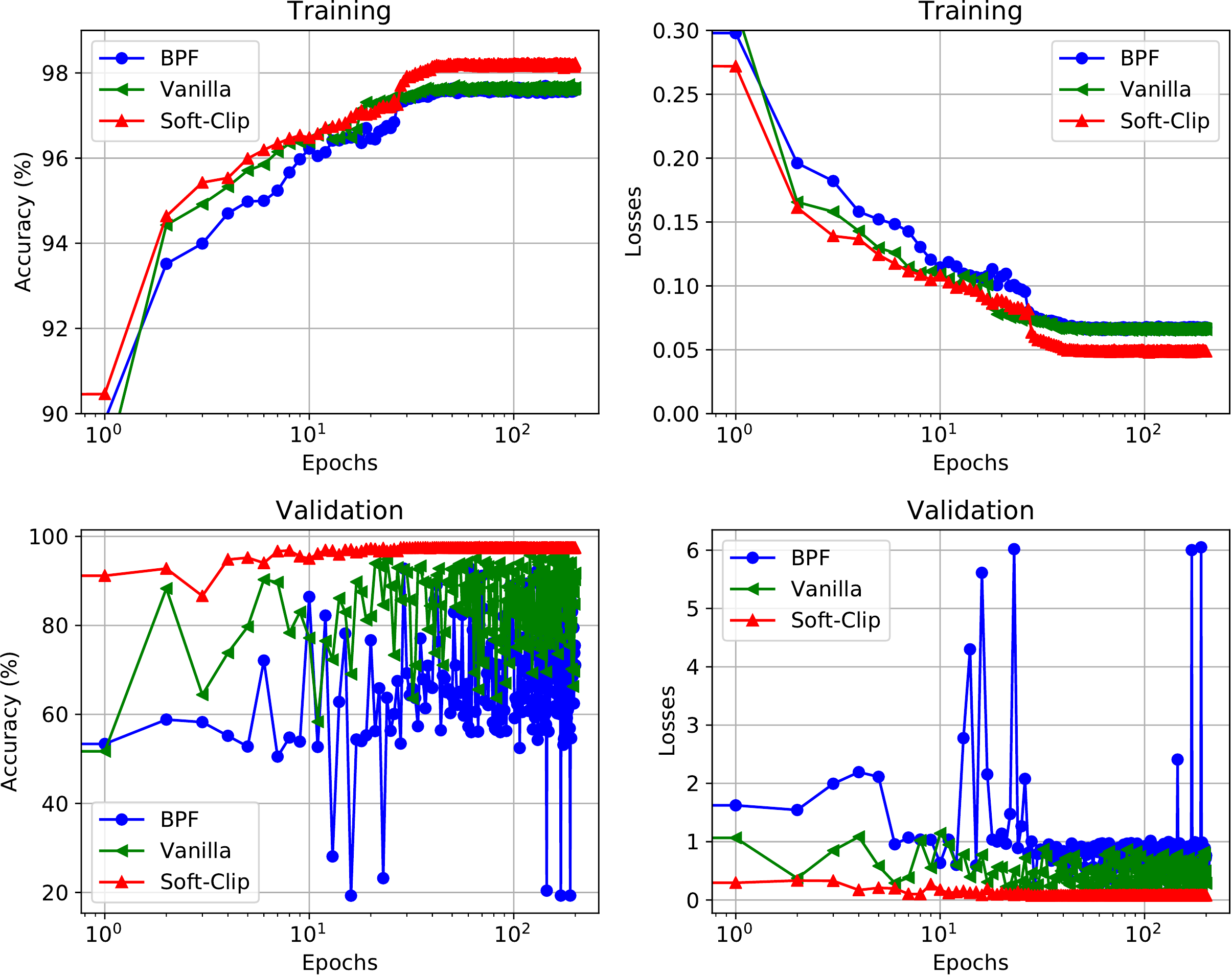}\llap{
        \parbox[b]{12.2cm}{(a)\hspace{6cm}(b)\\\rule{0ex}{4.8cm}(c)\hspace{6cm}(d)\\\rule{0ex}{4.5cm}}
    }
    \caption{Training process of 80\%--20\% chronological split with different preprocessing schemes: (a) Accuracy on Training Set, (b) Loss on Training Set, (c) Accuracy on Validation Set, (d) Loss on Validation Set. Soft-clip via logistic function in (c) is the most stable method and reaches highest validation accuracy.}
    \label{fig:preprocessing}
\end{figure}

\paragraph{Effect of soft-clipping}
%The convolutional neural networks trained on 40,000 samples are tested on 20,000 events to evaluate the accuracy after every epoch.
During the CNN training process, the network is tested after every epoch to evaluate its accuracy.
Figure~\ref{fig:preprocessing} shows the training loss, defined in equation~\eqref{eq:xentropy}, and testing accuracy, defined in equation~\eqref{eq:accuracy}, versus the number of epochs.
The proposed network with soft clipping (red) reaches 97\% accuracy after 40 epochs and becomes stable even though the training loss keeps going down.
On the other hand, without soft clipping (blue and green), the validation accuracy of the network slowly increases but exhibits a large oscillation centered around 80\% and 85\% accuracy, even though the training loss continues to decrease.
Thus with proper preprocessing, the trained CNN can reliably determine if a given 20-sec time window contains a P wave, S wave, or noise phase, and assess the likelihood of that decision.

\newpage
\section{Matched filter}
\label{sec:mf}

The analysis procedure of matched filter detection generally follows \cite{Meng2013} and is briefly described here.
Over 6,500 cataloged events between 2008/08/01 and 2008/08/30 are used to extract 6-sec templates.
A 2--8\,Hz band-pass filter is applied to enhance the strength of local earthquake signals, and the filtered waveforms are downsampled to 20\,Hz.
The 6-sec template window starts 1\,s before either the P wave on the vertical component or the S wave on horizontal components.
To avoid noisy traces, we measure the noise energy in a 6-sec window ahead of the template and define the corresponding signal-to-noise ratio (SNR) as the ratio between the energy of the template and noise energy.
Only traces with SNR above $5.0$ are used to cross-correlate with continuous data and output the cross-correlation (CC) function.
Stacked cross-correlation values on multiple stations are used to detect candidate events with a threshold of nine times the median absolute deviation (MAD) of the daily stacked correlation trace.
We select 2008/08/30 as the testing day since it has the most cataloged events, approximately 300.
Eventually, we end up with approximately 1,300 events and 12,200 phase picks that are detected on at least three stations.

\end{document}